\documentclass{aa}

\usepackage{graphicx}
\usepackage{lscape}
\usepackage{afterpage}

\usepackage{natbib}
\bibpunct{(}{)}{;}{a}{}{,} 

\begin{document}
\title{Studying the kinematic asymmetries of disks and post-coalescence mergers using a new `kinemetry' criterion}
   \author{Enrica Bellocchi\inst{1}, Santiago Arribas\inst{1}, Luis Colina\inst{1} }
   \institute{$^1$ Centro de Astrobiolog{\'i}a (CSIC-INTA),
     Ctra de Ajalvir km 4, 28850, Torrej{\'o}n de Ardoz, Madrid, Spain.\\
                   \email{bellochie@cab.inta-csic.es}}
\date{}                
\abstract{}
{(Ultra) Luminous Infrared Galaxies [(U)LIRGs] are important galaxy populations for studying galaxy evolution, and responsible for a significant fraction of the past star formation beyond z $\sim$ 1. Local (U)LIRGs can be used to study criteria adequate to characterize similar high-z populations. Of particular interest is to identify reliable kinematic-based methods capable of distinguishing {\it disks}{} and {\it mergers}, as their relative fraction is a key observational input to constrain different evolutionary scenarios.} 
{Our goal is to analyze in detail the kinematics of the ionized gas of a small sample of LIRGs and study criteria able to characterize the status of these systems.}
 {We have obtained VIMOS/VLT optical integral field spectroscopy (IFS) data for a sample of 4 LIRGs which have been selected at a similar distance ($\sim$ 70 Mpc) to avoid relative resolution effects. 
 They have been classified in two groups (\textit{isolated disk} and post-coalescence $mergers$) according to their morphology. The $kinemetry$ method (developed by Krajnovi{\'c} and coworkers) is used to characterize the kinematic properties of these galaxies and to discuss new criteria for distinguishing their status.}
 {We present and discuss new kinematic maps (i.e., velocity field and velocity dispersion) for these four galaxies. When the full 2-D kinematic information of the IFS data is analyzed with $kinemetry$, their morphological and kinematic classifications are consistent, with $disks$ having lower kinematic asymmetries than post-coalescence $mergers$. We propose and discuss a new kinematic criterion to differentiate these two groups. In particular, we introduce a weighting which favors the outer parts of the kinematic maps when computing the total asymmetries. This is motivated by the fact that post-coalescence mergers show relatively small kinematic asymmetries in the inner parts as a consequence of the rapid relaxation into a rotating disk, with the outer parts being still out of equilibrium (i.e., larger asymmetries). This new criterion distinguishes better these two categories and has the advantage of being less sensitive to angular resolution effects. According to the previous kinemetry-based work aimed at distinguishing disks and mergers at high-z, the present post-coalescence systems would have been classified as disks. This indicates that the separation of disks from mergers is subjective to the definition of `merger'. It also suggests that previous estimates of the merger/disk ratio could have been underestimated, but larger samples are necessary to establish a firmer conclusion. }

\keywords{galaxies -- kinematics -- luminous infrared galaxies -- integral field spectroscopy}

\titlerunning{Kinematic asymmetries of disks and post-coalescence mergers}
\authorrunning{Bellocchi et al.}

\maketitle

%
\section{Introduction}

The kinematic characterization of high-z galaxy populations is a key observational element to distinguish between different galaxy evolutionary scenarios, since it may help to find out the fraction of rotating disks and mergers at different cosmic epochs. This is  a key observational input to constrain the relative role of major mergers vs. steady cool gas accretion in shaping galaxies, which still is under discussion (e.g. \citealt{genzel01, T08, dekel09,  FS09, lemoine09, lemoine10, Bou11, Epi11}).

With the aid of integral field spectroscopy (IFS) we can resolve the kinematic status and internal processes at work within galaxies and better understand the role of the dominant mechanisms involved at early epochs. For instance, \cite{FS09} classified the SINS sample (i.e., 63 galaxies with $1.3< z<2.6$) concluding that these objects can be classified as: `rotation-dominated', interacting or merging systems and `dispersion-dominated' (i.e., dominated by large amounts of random motions). They found these types in a similar fraction (i.e., 1/3 each). They found a small fraction of galaxies showing signs of major merger events (i.e., \citealt{FS06, Law09, genzel06, genzel08}) therefore suggesting that the majority of the star forming galaxies at these redshifts is fed by gas via continuous cold flows along streams (including minor mergers). 

At intermediate redshift (z $\sim$ 0.6), spatially resolved kinematics has revealed a large fraction of chaotic velocity fields, supporting that most local spirals were rebuilt after a major merger since z = 1 (\citealt{hammer05, puech10}). At similar redshifts, \cite{Flores06} and \cite{Yang08} investigated the velocity fields of galaxies with IFS, and derived that only 35\% of their sources can be classified as rotating disks while the rest show perturbed or complex kinematics. 

Some local galaxy populations are particularly relevant for the study of galaxy evolution. This is the case of nearby Luminous and Ultraluminous Infrared Galaxies (LIRGs, L$_{IR}$ = [8-1000 $\mu$m] =10$^{11}$ - 10$^{12}$ L$_\odot$, and ULIRGs, L$_{IR} >$ 10$^{12}$ L$_\odot$, respectively). Despite these objects are relatively rare in the local universe, they are much more numerous at high-z and are responsible of a significant fraction of the whole past star formation beyond z $\sim$ 1 (e.g., \citealt{LF05, PG05, PPG08}). These objects exhibit a large  range of interacting/merging properties, from mostly isolated discs for low luminosity LIRGs (e.g. log L$_{IR}$ $\leq$ 11.3, \citealt{A04}) to a majority of merger remnants for ULIRGs (e.g. log L$_{IR}$  $\geq$ 12, \citealt{V02}). Local (U)LIRGs were initially considered as the local counterpart of high-z (U)LIRGs like those discovered by Spitzer and the more luminous submillimiter galaxies (SMGs; e.g. \citealt{B02, T06}). The study of local (U)LIRGs is also important to better understand the star formation history in the universe and it can provide an opportunity to correlate the properties of local and high-z populations. Moreover, large scale turbulences, tidal tails and outflows can be studied in detail allowing a good characterization of these objects. However, there are some discussion about the similarities between local and distant (U)LIRGs (e.g., \citealt{RJ11} and references therein).

Recently several authors have analyzed the velocity fields and velocity dispersion maps of different galaxy samples with the aim of discerning merging and non-merging systems based on their kinematic properties using the $kinemetry$ methodology (\citealt{K06}). \cite{jesseit07} and \cite{kron07} have investigated the distortions in the velocity fields of simulated interacting $disc/merger$ galaxies at different redshifts in the range $z = 0-1$, while \cite{S08} (hereafter, S08) provide a straightforward mean of classifying $z\sim2$ galaxies, observed with SINFONI/VLT, as $disks$ or $mergers$ based on the asymmetries of their velocity field and velocity dispersion maps. They compare and distinguish two kinematic classes characterized by recent major merger events ({i.e., \it mergers}) and those without signs of interacting or recent merger activity (i.e., {\it disks}).

Similar works based on IFS data (i.e., OSIRIS/Keck, \citealt{BZ09, Gon10}) are focused on nearby objects (i.e., supercompact UV-Luminous Galaxies (ScUVLGs) and Lyman Break Analogs (LBAs)). The morphology and kinematics of LBAs projected at high redshift show similarities with compact and dispersion-dominated $z\sim2$ galaxies (\citealt{BZ09}). This suggests that galaxy interaction and/or mergers could also be driving the dynamics of dispersion-dominated $z\sim2$ galaxies.

In this paper we present new spatially resolved kinematics of four local LIRGs obtained with the VLT/VIMOS IFU as part of a larger project to characterize the properties of (U)LIRGs on the basis of optical and infrared Integral Field Spectroscopy (\citealt{Co05, A08, MI10, RZ11} and references therein). The fact that two of our LIRGs are post-coalescence mergers will offer the opportunity to explore the potential of the $kinemetry$ method when analyzing the velocity fields and velocity dispersion maps in moderately disturbed and partially relaxed systems. The paper is structured as follows. In Section 2, we present the sample selection as well as some  details about the observations, data reductions, line fitting and velocity map construction. Section 3 is devoted to the description of the main kinematic properties of the individual objects. In Section 4, the $kinemetry$ analysis and its potential to distinguish $discs$ and $mergers$ are presented and discussed with the aid of different asymmetry planes. Finally, the main results and conclusions are summarized in Section 5. 

Throughout the paper we will consider H$_0$ = 70 km/s/Mpc, $\Omega_M$ = 0.3 and $\Omega_\Lambda$ = 0.7.

\section{The VIMOS sample, observations, data reduction, line fitting and map construction}

\subsection{The sample and morphological classification}

For the present analysis we have selected 4 LIRG galaxies at a similar distance ($\sim$ 70 Mpc): two of them (i.e., IRAS F11255-4120, IRAS F10567-4310) are morphologically classified as {\it class 0} objects and two (i.e., IRAS F04315-0840, IRAS F21453-3511) as {\it class 2} objects according to \cite{A08} and \cite{RZ10} (hereafter, Paper I and Paper III respectively). In this classification scheme {\it class 0} is defined as an object with relatively symmetric morphology, appearing to be isolated with no evidence for strong past or ongoing interaction (hereafter, $disks$), while {\it class 2} is an object with a morphology suggesting a post-coalescence merging phase, with a single nucleus or two distinct nuclei at a projected distance D $<$ 1.5 kpc (hereafter, {\it post-coalescence mergers}). We have chosen these four systems at the same distance as they will allow us to discuss their relative kinematic properties avoiding linear resolution dependency effects. In table \ref{table_sample} we present the main properties of the sample: note that the systems classified as post-coalescence mergers have higher L$_{IR}$ than the disks.

\begin{table*}
\centering
\caption{General properties of the LIRGs sub-sample.}
\label{table_sample}
\begin{scriptsize}
\begin{tabular}{c c c c c c c c c  c c} \\
\hline\hline\noalign{\smallskip}  
ID1 &  ID2 &  $\alpha$ & $\delta$  &  $z$ &  D&  scale & log L$_{IR}$  & Class  &  References  \\
IRAS & Other & (J2000)& (J2000) &  &  (Mpc) &  (pc/arcsec)     & (L$_{\odot}$)    &     &   \\ 
(1) & (2) &  (3)& (4) &(5)&(6) & (7)  & (8)  & (9) & (10)    \\
\hline\noalign{\smallskip} 	
 F11255-4120   &   ESO 319-G022 & 11:27:54.1 & -41:36:52 &  0.016351 & 70.9 & 333 & 11.04  &  0 & 1      \\
 F10567-4310   &  ESO 264-G057 &  10:59:01.8 &  -43:26:26 &  0.017199 &  74.6  &  350 & 11.07  &  0 &  1    \\
 F04315-0840   &  NGC 1614 & 04:33:59.8 &  -08:34:44 & 0.015983 & 69.1  & 325 & 11.69  &   2 & 1, 2   \\
 F21453-3511   &  NGC 7130 & 21:48:19.5 &  -34:57:05 & 0.016151 & 70.0 & 329 & 11.41  &   2 & 1, 2    \\
\hline\hline
\end{tabular}
\end{scriptsize}
  \begin{minipage}[h]{16.8cm}
   \vskip0.1cm\hskip0.0cm
\footnotesize
\tablefoot{ Col (1): object designation in the IRAS Faint source catalogue (FSC). Col (2): other name. Col (3) and (4): right ascension (hours, minutes and seconds) and declination (degrees, arcminutes and arcseconds) from the IRAS FSC. Col (5): redshift of the IRAS sources from the NASA Extragalactic Database (NED). Col (6): luminosity distances assuming a $\Lambda$DCM cosmology with H$_0$ = 70 km/s/Mpc, $\Omega_M$ = 0.3 and $\Omega_\Lambda$ = 0.7, using the Edward L. Wright Cosmology calculator. Col (7): scale. Col (8): infrared luminosity (L$_{IR}$= L(8-1000) $\mu$m), in units of solar bolometric luminosity, calculated using the fluxes in the four IRAS bands as given in \cite{sanders03} when available. Otherwise the standard prescription in \cite{SM96} with the values in the IRAS Point and Faint source catalogues was used. Col (9): Morphology class.  For further information see table 1 in \cite{RZ10}.}
\tablebib{(1) \cite{V95}; (2) \cite{C03}. }
	\end{minipage}
\end{table*}

\subsection{Observations}

The observations were carried out in service mode using the Integral Field Unit of VIMOS (\citealt{Lfevre03}), at the Very Large Telescope (VLT), covering the spectral range $(5250-7400)$ \AA\ with the high resolution mode `HR-orange' (grating GG435) and a mean spectral resolution of 3470. The field of view (FoV) in this configuration is 27$^{\prime\prime}$ $\times$ 27$^{\prime\prime}$, with a spaxel scale of 0.67$^{\prime\prime}$ per fiber (i.e., 1600 spectra are obtained simultaneously from 40 $\times$ 40  fibers array).  A square 4 pointing dithering pattern is used, with a relative offset of 2.7$^{\prime\prime}$ (i.e., 4 spaxels). The exposure time per pointing ranges from 720 to 850 seconds, such that, the total integration time per galaxy is between 2880 and 3400 seconds. For further details about the observations see Paper I.

\subsection{Data reduction}
\label{reduc}

The VIMOS data are reduced with a combination of the pipeline {\it Esorex} (version 3.5.1 and 3.6.5) included in the pipeline provided by ESO, and different customized IDL and IRAF scripts.
The basic data reduction is performed using the {\it Esorex} pipeline (bias subtraction, flat field correction, spectra tracing and extraction, correction of fiber and pixel transmission and relative flux calibration). Then four quadrants per pointing are reduced individually and combined in a single data cube associated for each pointing. After that, the four independent dithered pointings are combined together to end up with the final `super-cube', containing 44 $\times$ 44 spaxels for each object (i.e., 1936 spectra). Further details about the data reduction for the whole sample can be found in Paper I and in Paper III.

The wavelength calibration, the instrumental profile and fiber-to-fiber transmission correction are checked for the four galaxies using the [OI]$\lambda$6300.3 \AA\ sky line as in Paper I. The results is focused on the H$\alpha$ and [NII]$\lambda\lambda$6548.1, 6583.4 \AA\ emission lines, so that the sky line here considered is suitable for its proximity to these lines. In table \ref{calibration} the average values for the central wavelength $<\lambda_c>$ and (instrumental) width  $<\sigma_{INS}>$, with their standard deviations, are shown for the four galaxies.

\begin{table}[h]
\centering
\caption{Results of the calibration check using the [OI]$\lambda$6300.3 \AA\ sky line. }
\label{calibration}
\begin{small}
\begin{tabular}{ l c c c } 
\hline\hline\noalign{\smallskip} 
 IRAS Galaxy &  $<\lambda_c> \pm \Delta\lambda_c$ [\AA] &  $<\sigma_{INS}>\pm std$ [\AA]  \\
\hline\hline\noalign{\smallskip} 
 F11255-4120 &  6300.42 $\pm$ 0.15  & 0.78 $\pm$ 0.09       \\
\hline\noalign{\smallskip} 
  F10567-4310 &  6300.17 $\pm$ 0.13  &  0.77 $\pm$ 0.09     \\
\hline\noalign{\smallskip} 
  F04315-0840 &  6300.24 $\pm$ 0.10 & 0.78 $\pm$ 0.08    \\
\hline\noalign{\smallskip} 	
  F21453-3511  &  6300.27 $\pm$ 0.11  & 0.77 $\pm$ 0.06     \\
\hline
\end{tabular}
\end{small}
\begin{minipage}[h]{9cm}
	\vskip0.2cm\hskip0.0cm
\footnotesize
\tablefoot{Typical values of the central wavelength and width distribution of the [OI]$\lambda$6300.3 \AA\ sky line registered in the data cubes along with their standard deviations (errors). }
\end{minipage}

\end{table}

\subsection{Line fitting and map construction}

As mentioned in the previous paragraph, we are interested in the H$\alpha$ - [NII] emission line complex analysis. The observed emission profiles of the individual spectra are fitted to a Gaussian function using an IDL routine (i.e., MPFITEXPR, implemented by C. B. Markwardt). This algorithm finds the best set of model parameters which match the data and it is able to fix the line intensity ratios  and the wavelength differences according to the atomic parameters when adjusting multiple emission lines (i.e., the H$\alpha$ - [NII] complex). The MPFIT routine provides errors for the output parameters and gives an estimate of the goodness of the fit.
For each emission line we end up with the following information: central wavelength ($\lambda_c$), intrinsic width ($\sigma_{line})$ (i.e., $\sigma_{line} = \sqrt{\sigma_{obs}^2 - \sigma_{INS}^2}$) and flux intensity along with their respective errors. To obtain the radial velocity uncertainty estimates, wavelength calibration and fitting errors were combined in quadrature, giving a global wavelength error of  $ \Delta\lambda_{tot} = \sqrt{\Delta\lambda_{MPFIT}^2 +  \Delta\lambda_{c}^2}$. For high S/N spectra the wavelength calibration error can be considered the main source of uncertainty since the fitting errors are typically substantially smaller (and vice versa for low S/N spectra). For the fitting errors, we derive values generally smaller than 0.2 \AA\ while the wavelength calibration errors are of the order of $\sim$ 0.12 \AA\, as shown in Tab. \ref{calibration}. An average value of the $\sigma_{INS}$ of (85 $\pm$ 9) km/s is derived at the corresponding wavelength, along with a mean FWHM of (1.82 $\pm$ 0.19) \AA\ {}.

We start fitting all the lines to single Gaussian profiles since the main component usually extends over the entire galaxy. Two-Gaussian profiles (i.e., main and broad components) are needed for the inner regions of the four objects. Using a collection of procedures written in IDL code (i.e, {\bf jmaplot}, \citealt{MA04}), we generate flux intensity, velocity field and velocity dispersion maps, respectively, for the main and broad components (Figures \ref{class0_1} - \ref{class2}) along with a VIMOS continuum image. When HST images are available (i.e, NICMOS F160W (H-band) and ACS F814W (I band)) they are also shown in the panels. 

\section {Kinematic properties of individual objects}

In this section we describe the global kinematic behavior as inferred from our IFS maps (i.e, flux intensity, velocity field  and velocity dispersion). We will also infer some kinematic parameters (e.g., $\sigma_c$, v$_c/\sigma_c$, v$_{shear}/\Sigma$ and dynamical mass M$_{dyn}$) useful to characterize these systems.

The ratio of the maximum circular rotation velocity v$_c$ (i.e., the half of the observed peak-to-peak velocity) and the central velocity dispersion $\sigma_c$ measures the nature of the gravitational support of a system in equilibrium. A v$_c$/$\sigma_c$ $\geq$ 1 is the signature of a rotation-dominated system while a lower value (i.e., v$_c$/$\sigma_c <$ 1) means that the object is dispersion-dominated, as in the case of elliptical galaxies, where the galaxy is sustained against gravitational collapse by the pressure originated by random motions of the stars. When analyzing the v$_c/\sigma_c$ parameter it is important to realize that the presence of flows, superwinds or other merger-induced processes are factors that may lead to large v$_c$ and/or $\sigma_c$ values. Therefore we compute the circular velocity v$_c$ and the central velocity dispersion $\sigma_c$ using the component of one-Gaussian model fit (i.e., 1c) and the main (i.e., the one defining the systemic behaviour) component of the two-Gaussian model fit (i.e., 2c).

Since class 0 objects show the features of an {\it ideal rotating disk} \footnote{ We refer to {\it ideal rotating disk} as a thin disk with gas clouds kinematically characterized by: a velocity field that peaks at the galaxy major axis and goes to zero along the minor axis  of each orbit, while the velocity dispersion will be constant long each orbit and will decrease between orbits of increasing major axes (see description in Sec. 4.1)} we correct their kinematic values for their respective inclination; class 2 objects show the velocity fields distorted such that we derive their kinematic values with and without correcting for the inclination.

The ratio between the velocity shear v$_{shear}$ and the global velocity dispersion in the whole galaxy $\Sigma$ has been derived too, as done in \cite{Gon10}. The v$_{shear}$ (not corrected for the inclination of the object) has been computed as the median of the 5-percentile at each end of the velocity distribution, for the v$_{max}$ and v$_{min}$ and then defined as v$_{shear}$ = $\frac{1}{2} (v_{max} - v_{min}$).

We have also used the kinematic information to derive the dynamical mass $M_{dyn}$. This mass takes into account the whole gravitational field and so dark matter, gas and stellar components are included. Spiral galaxies can be modeled using a flat component for the disk and a spheroidal component for the central concentration (i.e., a de Vaucouleurs profile and a massive halo). Assuming that the central regions are virialized, the observed velocity dispersion of the ionized gas $\sigma_c$ can provide a good estimate for the disk's dynamical mass ($M_{dyn}$, e.g., \citealt{Co05}). We use the following relation:

\begin{equation}
\hspace{1cm}M_{dyn} = m \hspace{1mm} 10^6 \hspace{1mm}R_{hm}\hspace{1mm}\sigma_c^2\hspace{1mm} M_{\odot}
\end{equation}

where $R_{hm}$ is the half-mass radius in $kpc$ and $\sigma_c$ is the central velocity dispersion in $km/s$. The $R_{hm}$ parameter is not an observable itself and therefore it cannot be measured directly, so it can be inferred from measuring the radius that encloses half luminosity (i.e., $R_{eff}$). The R$_{eff}$ were obtained from existing 2MASS near-infrared imaging. The extinction in the near-IR is a factor of 5-10 smaller than that in the optical and a proper way to minimize these effects is to measure the half-light radius in the $H$ or $K$ bands as done here. The parameter $m$ depends on the mass distribution: its value ranges from 1.4, for a King stellar mass distribution with a tidal-to-core ratio of 50, which is  a good representation of elliptical galaxies (\citealt{T02}), to 1.75 for a polytropic sphere with a density index covering a range of values (\citealt{Spitzer87}), to 2.1 for a de Vaucouleurs mass distribution (\citealt{Combes95}). As in \cite{Co05} we will assume $m =1.75$. All these results are shown in the table \ref{pixel}; their inclinations are drawn from the NED site and are consistent with our own estimates from the H$\alpha$ distribution.

\subsection  {\bf IRAS F11255-4120 (ESO 319-G022) }

This is a class 0 object, appearing to be a single isolated galaxy with a relatively symmetric morphology according to its DSS image. It is a barred spiral with a ring extended up to $\sim$ 4  kpc from the nuclear region, clearly detected in our H$\alpha$ image. The orientation of the bar is different in the VIMOS continuum image (P.A. $\sim$ 110$^{\circ}$) and in the H$\alpha$ image (P.A. $\sim$ 150$^{\circ}$).

\begin{figure}
\includegraphics[width=0.48\textwidth]{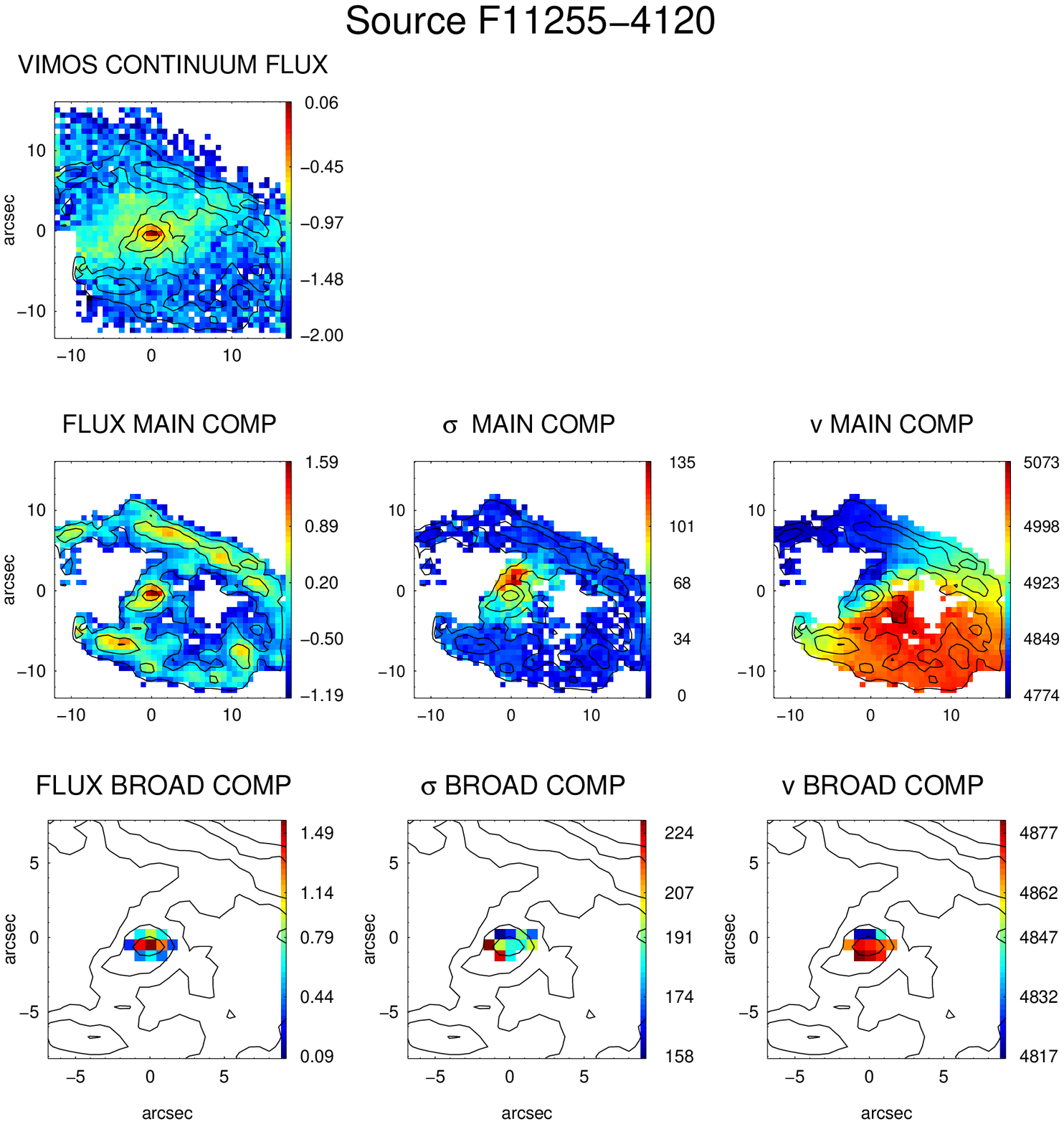}
\includegraphics[width=0.48\textwidth]{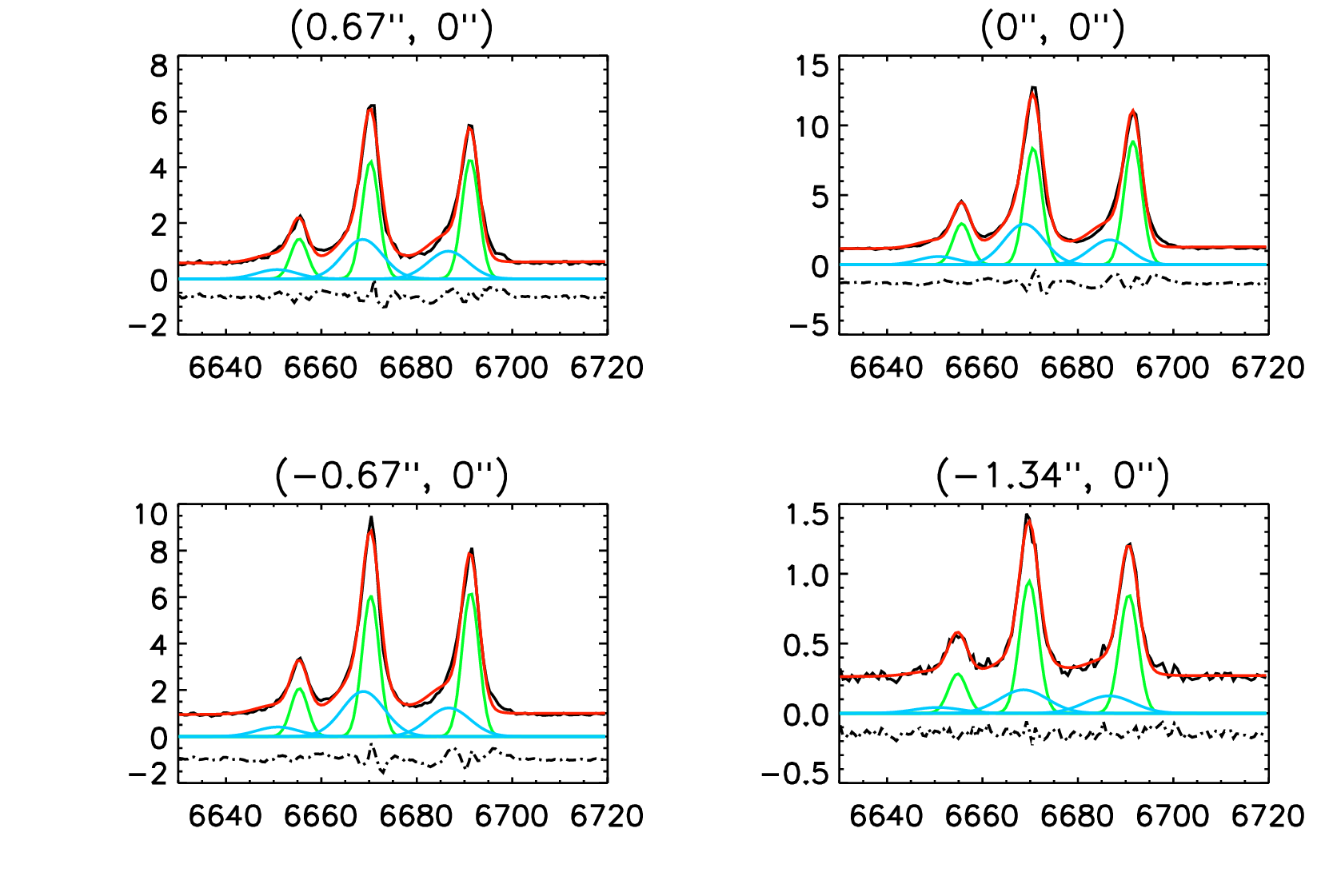}
\caption{\small {\bf Top panel:}  H$\alpha$ maps for the class 0 object IRAS F11255-4120. VIMOS continuum image; HST continuum images are not available for this galaxy. {\it Middle row:} the flux intensity, velocity dispersion $\sigma$ (km/s) and velocity field $v$ maps (km/s) for the main component. {\it On the bottom:} respective maps for the broad component. Note the different FoV between the main and the broad components. The latter has been zoomed in since it covers a small area. The flux intensity maps are represented in logarithmic scale and in arbitrary flux units. All the images are centered using the H$\alpha$ flux intensity peak and the iso-countors of the H$\alpha$ flux are overplotted. {\bf Bottom panel:} H$\alpha$-[NII] observed spectra of IRAS F11255-4120 for selected inner regions (indicated by the coordinates in the top label using the same reference system as in the top panel) where the main and broad component coexist. The red curve shows the total H$\alpha$-[NII] components as obtained from multi-components Gaussian fits. The green and blue curves represent respectively the main and broad components.}
\label{class0_1}
\end{figure}

For the main component, as shown in the top panel of Fig. \ref{class0_1}, the velocity dispersion ($\sigma$) map has a (almost) centrally peaked structure with values of 120 km/s (high values also in the bar structure) and a lower mean value ($\sim$ 20 km/s) is found in the ring. The location of this peak does not agree with those of the H$\alpha$ flux intensity and continuum maps with an offset of about 0.6 kpc (i.e., $\sim$ 2 arcsec). The kinematic center of the velocity field well agrees with the H$\alpha$ flux peak. The projection of the rotation axis (minor kinematic axis) seems to be shifted by about 30$^{\circ}$ with respect to the orientation of the bar in the H$\alpha$ flux intensity map and in agreement with the bar of the continuum. The observed velocity amplitude for the main component at a distance of 4 kpc from the H$_\alpha$ peak is of the order of (300 $\pm$ 8) km/s.

A small region of about $\sim$ 0.67 kpc x 0.89 kpc  in the nuclear region  shows traces of a second component, indicating the presence of a non-rotational mechanisms (e.g., outflow / wind). This component is blueshifted by 80 km/s with respect to the main one and shows a broad velocity dispersion $\sigma$ in the range (160 - 220) km/s and a velocity amplitude of the order of $(60 \pm 7)$ km/s.

The v$_c/\sigma_c$ and v$_{shear}/\Sigma$ have values of 2.3 and 2.7 respectively, supporting the idea that this galaxy is rotation-dominated: the kinematic properties of this galaxy seem to be consistent with its morphology (i.e., a rotating disk).

\subsection  {\bf{IRAS F10567-4310 (ESO 264-G057)} }

\begin{figure}
\begin{center}
\includegraphics[width=0.48\textwidth]{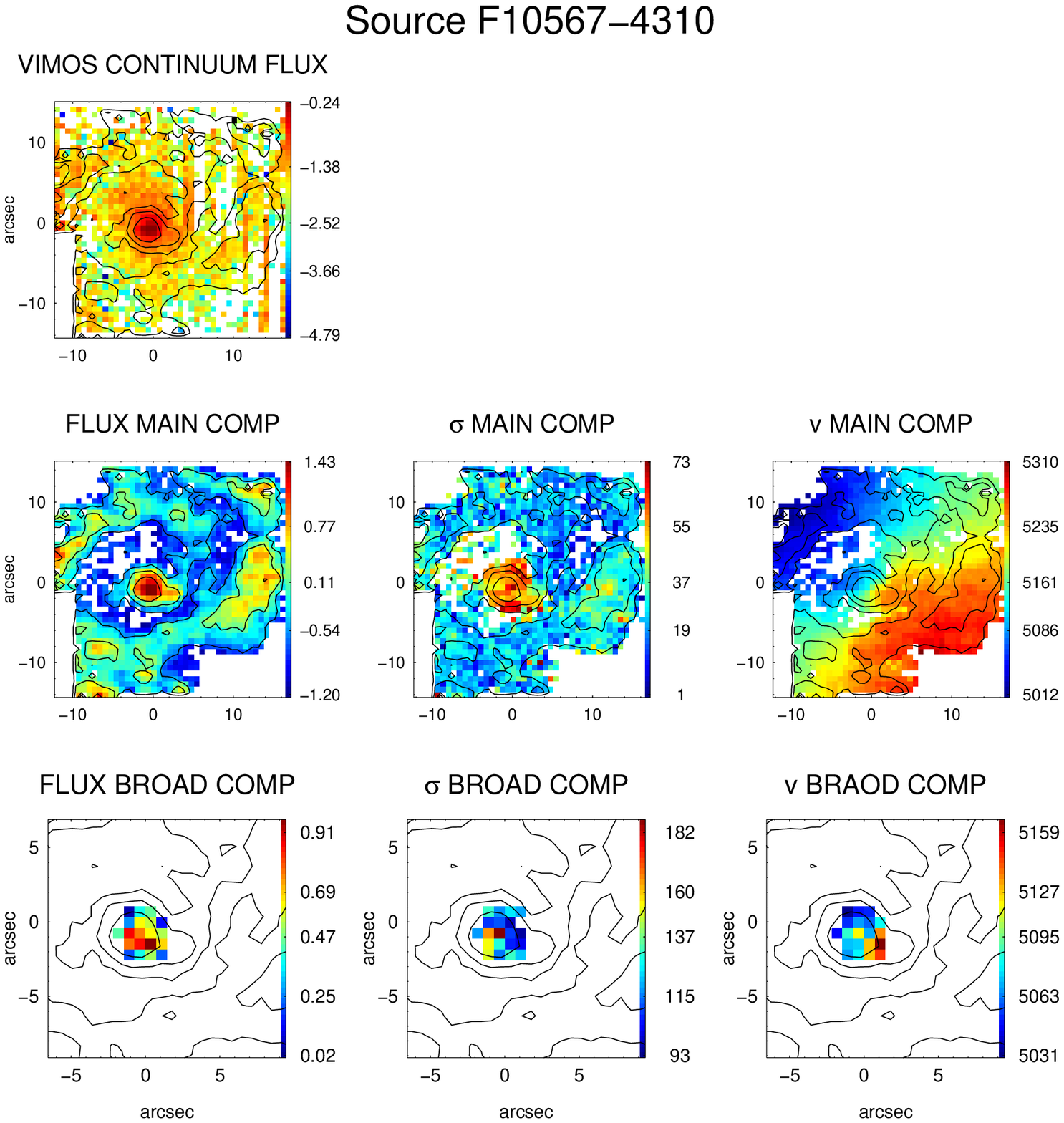}
\includegraphics[width=0.48\textwidth]{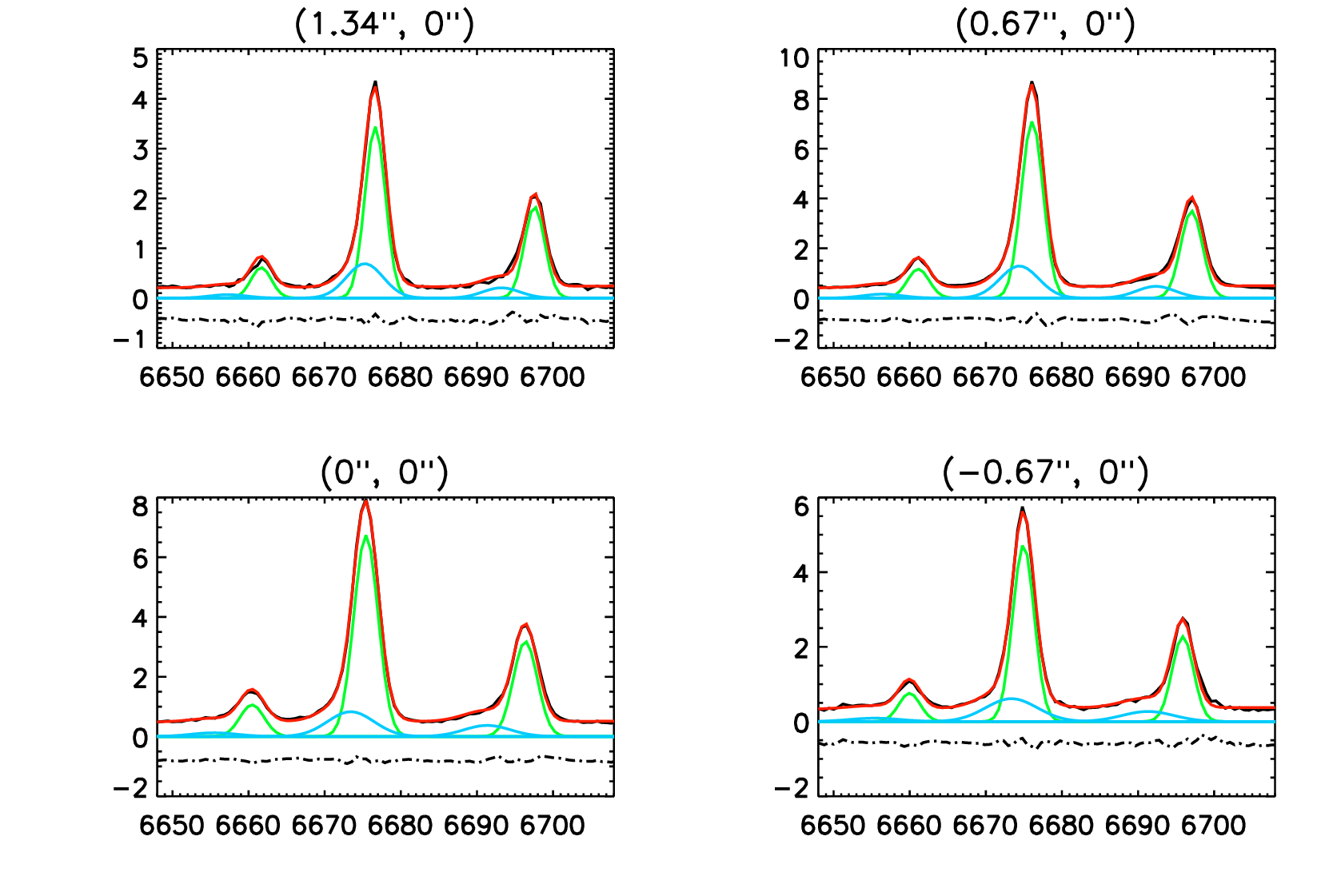}
\caption{\small {\bf Top panel:} H$\alpha$ maps for the class 0 object IRAS F10567-4310. VIMOS continuum image; HST continuum images are not available for this galaxy. {\it Middle row:} the flux intensity, velocity dispersion $\sigma$ (km/s) and velocity field $v$ maps (km/s) for the main component. {\it On the bottom:} respective maps for the broad component. Note the different FoV between the main and the broad components. The latter has been zoomed in since it covers a small area. The flux intensity maps are represented in logarithmic scale and in arbitrary flux units. All the images are centered using the H$_{\alpha}$ flux intensity peak and the iso-countors of the H$_{\alpha}$ flux are overplotted. {\bf Bottom panel:} H$_\alpha$-[NII] observed spectra of {IRAS F10567-4310} for selected inner regions (indicated by the coordinates in the top label using the same reference system as in the top panel) where the main and broad component coexist. The red curve shows the total H$\alpha$-[NII] components as obtained from multi-components Gaussian fits. The green and blue curves represent respectively the main and broad components.}
\label{class0}
\end{center}
\end{figure}

From the DSS image IRAS F10567-4310 is morphologically classified as class 0 object (see Paper III). The H$\alpha$ flux map inferred from our IFU data (top panel of Fig. \ref{class0}) shows a ring with a patchy distribution and its peak is in good positional agreement with the center of the of galaxy as inferred from the continuum.

The main component shows a centrally peaked velocity dispersion map with typical values of 60 - 70 km/s. Its general morphology resembles that of the H$\alpha$ flux map (i.e., regions with high H$\alpha$ surface brightness tend to have large velocity dispersion, with a mean value of 40 km/s). This component shows a disk-like regular velocity field with a clear rotation pattern, with the kinematic center in a good positional agreement with the H$\alpha$ flux peak. The observed velocity amplitude is of $(300 \pm 6)$ km/s computed at a distance of 4.7 kpc from the kinematic center. 

A secondary kinematic distinct component is identified in the nuclear region over an extension of $\sim$ 1.2 kpc x 0.9 kpc; it shows a velocity amplitude of (128 $\pm$ 9) km/s and it is blue-shifted by $\sim$ 70 km/s with respect to the main one; its velocity dispersion is in the range (90 - 180) km/s. 

The v$_c/\sigma_c$ parameter clearly classifies this object as rotation-dominated with v$_c/\sigma_c$ $\sim$ 4.6 and v$_{shear}/\Sigma \sim$ 3. Kinematics of this class 0 object is also consistent with its morphology.

\subsection  {\bf IRAS F04315-0840 (NGC 1614)}

\begin{figure}[!h]
\begin{center}
\includegraphics[width=0.48\textwidth]{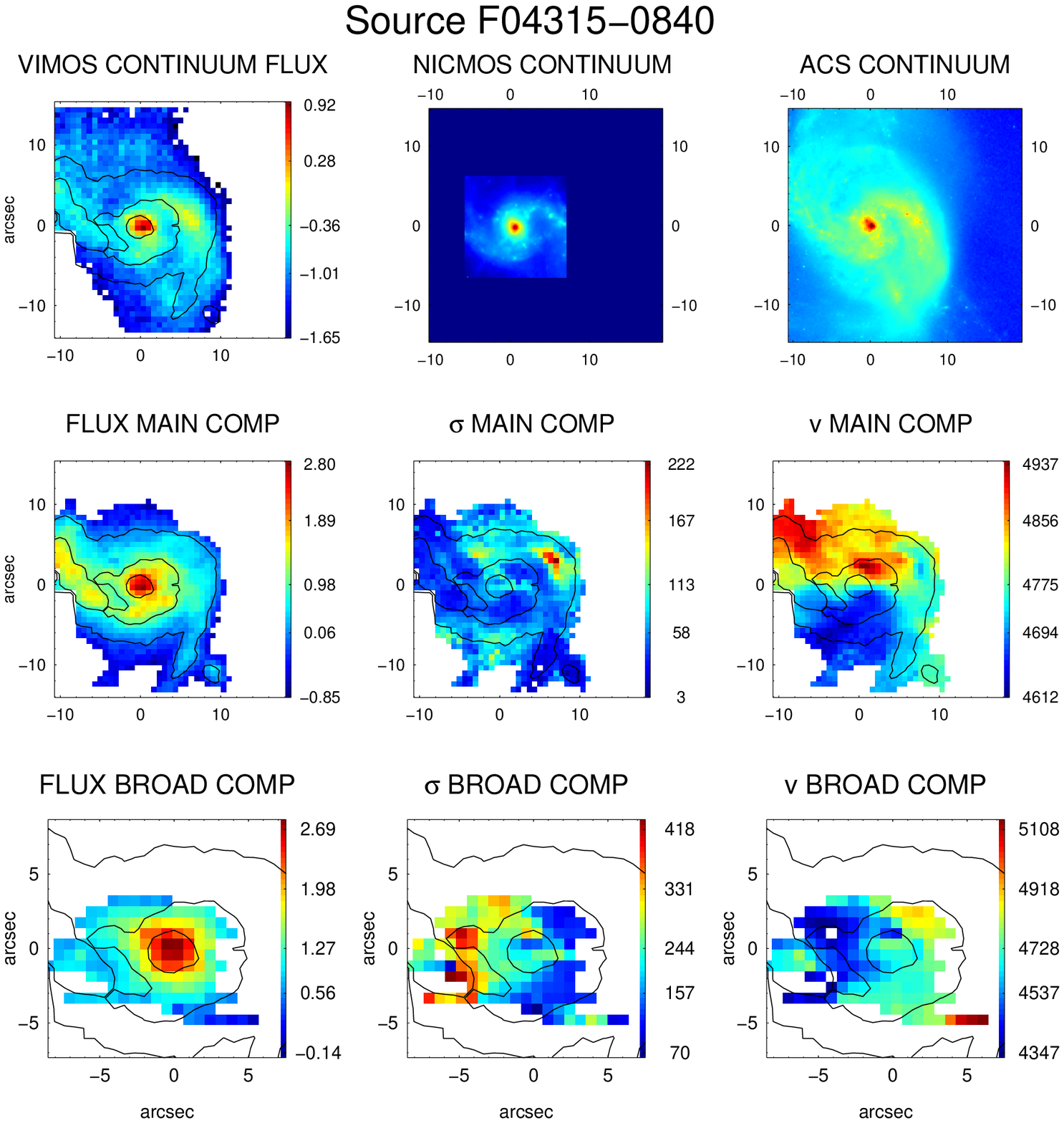}
\includegraphics[width=0.48\textwidth]{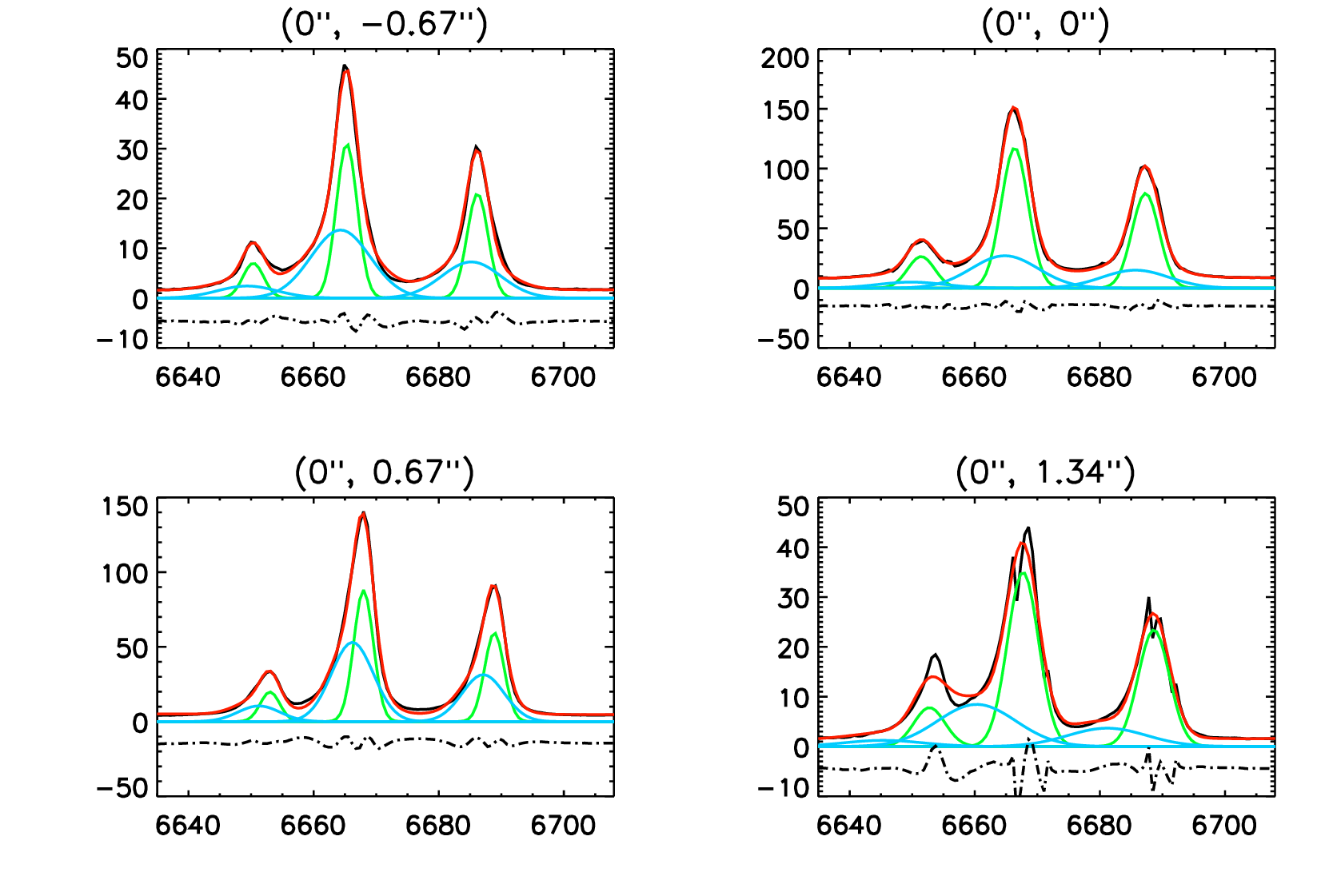}
\caption{\small {\bf Top panel:} H$\alpha$ maps for the class 2 object IRAS F04315-0840. VIMOS continuum image and HST continuum images (i.e., H and I bands). {\it Middle row:} flux intensity, velocity dispersion $\sigma$ (km/s) and velocity field $v$ maps (km/s) for the main component. {\it On the bottom:} respective maps for the broad component. Note the different FoV between the main and the broad components. The latter has been zoomed in since it covers a small area. The flux intensity maps are represented in logarithmic scale and in arbitrary flux units. All the images are centered using the H$\alpha$ peak and the iso-countors of the H$\alpha$ flux are overplotted. {\bf Bottom panel:} H$\alpha$-[NII] observed spectra of {IRAS F04315-0840} for selected inner regions (indicated in the top label using the same reference system as in the top panel) where the main and broad component coexist. The red curve shows the total H$\alpha$-[NII] components as obtained from multi-components Gaussian fits. The green and blue curves represent respectively the main and broad components.}
\label{class2_2}
\end{center}
\end{figure}

This is a class 2 object (Fig. \ref{class2_2}) with a relatively asymmetric morphology: the DSS image shows a tidal tail extending for 13 kpc from the nuclear region. This is a well studied, post-coalescence late merger, with a bright, spiral structure at scale of few kpc (1 - 3 kpc). The spiral structure in H$\alpha$ flux map shows a different orientation with respect to the continuum.

The velocity dispersion map of the main component has an irregular structure. Its peak (i.e., 220 km/s) is found at 2.4 kpc from the nucleus (i.e., H$\alpha$ flux peak) in the northern arm. The velocity field of the main component is somewhat distorted and chaotic with an amplitude of $(325 \pm 5)$ km/s. The projection of the rotation axis in the outer part (connecting the N-E to S-W part) is not aligned with that of the inner part (aligned as N-S direction).

A broad component is found in the inner region and covers a relatively large area of $\sim$ 2.4 kpc x 2.7 kpc. This component is blue-shifted by 50 km/s with respect to the main component. Its velocity dispersion is in the range (70 - 400) km/s with a velocity amplitude of (760 $\pm$ 13) km/s. The fact that the projection of the kinematic axis of this broad component has a shift of almost 90$^{\circ}$ with respect to the main component and the blue-shifted region of the velocity field shows the largest velocity dispersion (i.e., $\sim$ 400 km/s) supports the hypothesis of a dusty outflow, where the receding components (which are behind the disk) are obscured, making the whole profile relatively narrow with respect to the approaching component. 

We derive a $v_{c}/\sigma_c$ $>$ 1 even when the inclination correction is not included (i.e., $v_{c}/\sigma_c$ $\approx$ 1.4 - 2.3): it shows the dominance of an intrinsic rotation over random motions. Therefore, this parameter would classify this object as rotation dominated.

\subsection  {\bf IRAS F21453-3511 (NGC 7130)}

{IRAS F21453-3511} is a peculiar class 2 object, with traces of spiral and asymmetric morphology from the HST and DSS images. The ionized gas, as traced by the H$\alpha$ emission, is concentrated in the nuclear region and mostly in the northern spiral arm. 

The asymmetric velocity dispersion map of the main component shows higher values in the northern arm and its central part  (i.e., two local maxima can be revealed) with values of 80 km/s. The velocity field of the same component shows an asymmetric behavior where three main regions can be identified: the SW part having a mean value of 4900 km/s, the NW one with 4850 km/s and the E part with 4780 km/s. The NW part could have different inclination than the rest of the observed regions, so explaining this peculiar velocity structure. The kinematic center can be well identified with the H$_\alpha$ flux peak although the rotational axis it is not well defined. The photometric major axis in the NICMOS image seems to be orientated in the N-S direction and the ACS image seems to reveal a ring of knots with a major axis orientated along P.A. $\sim$ 135$^{\circ}$.

\begin{figure}
\begin{center}
\includegraphics[width=0.48\textwidth]{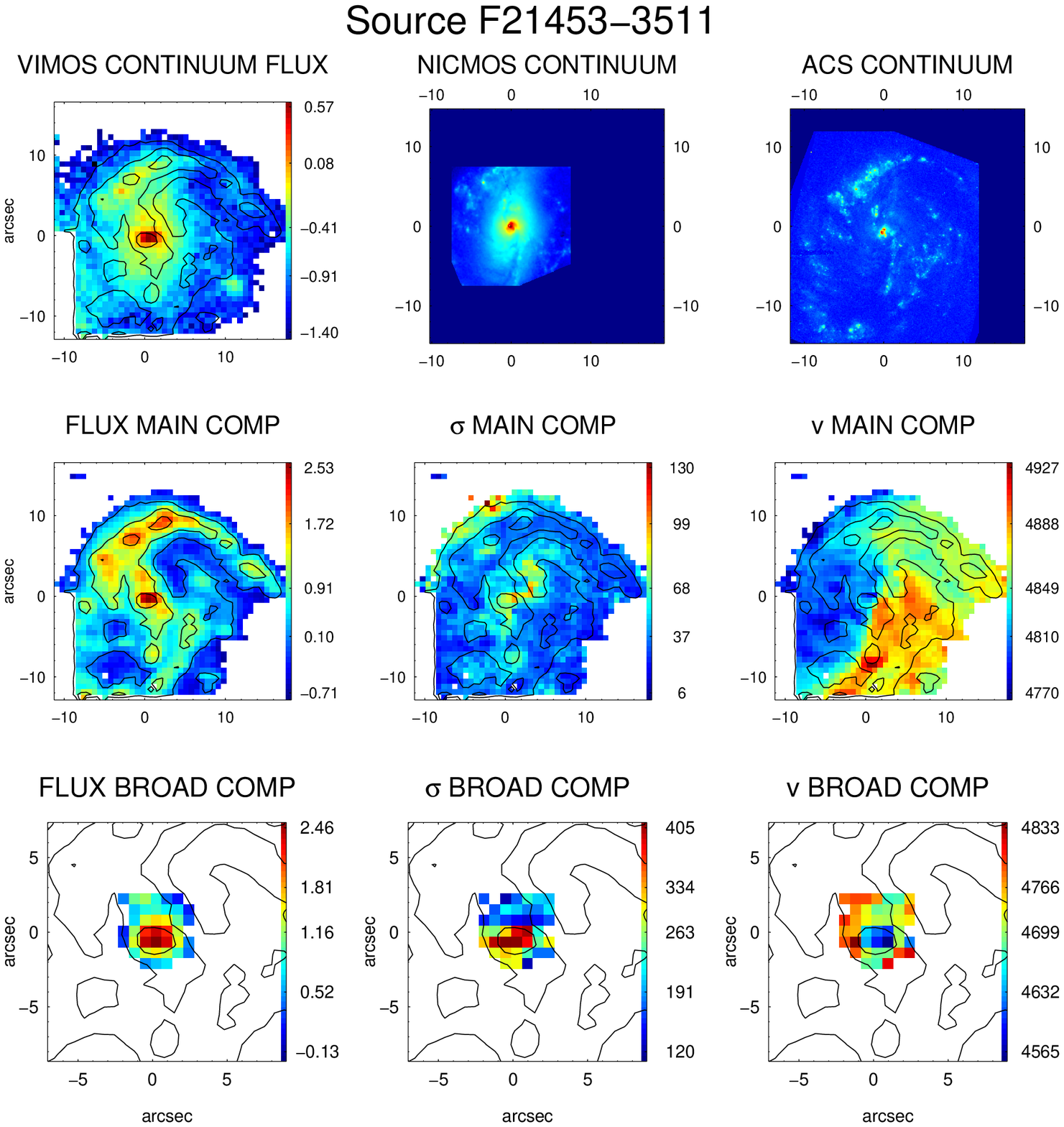}
\includegraphics[width=0.48\textwidth]{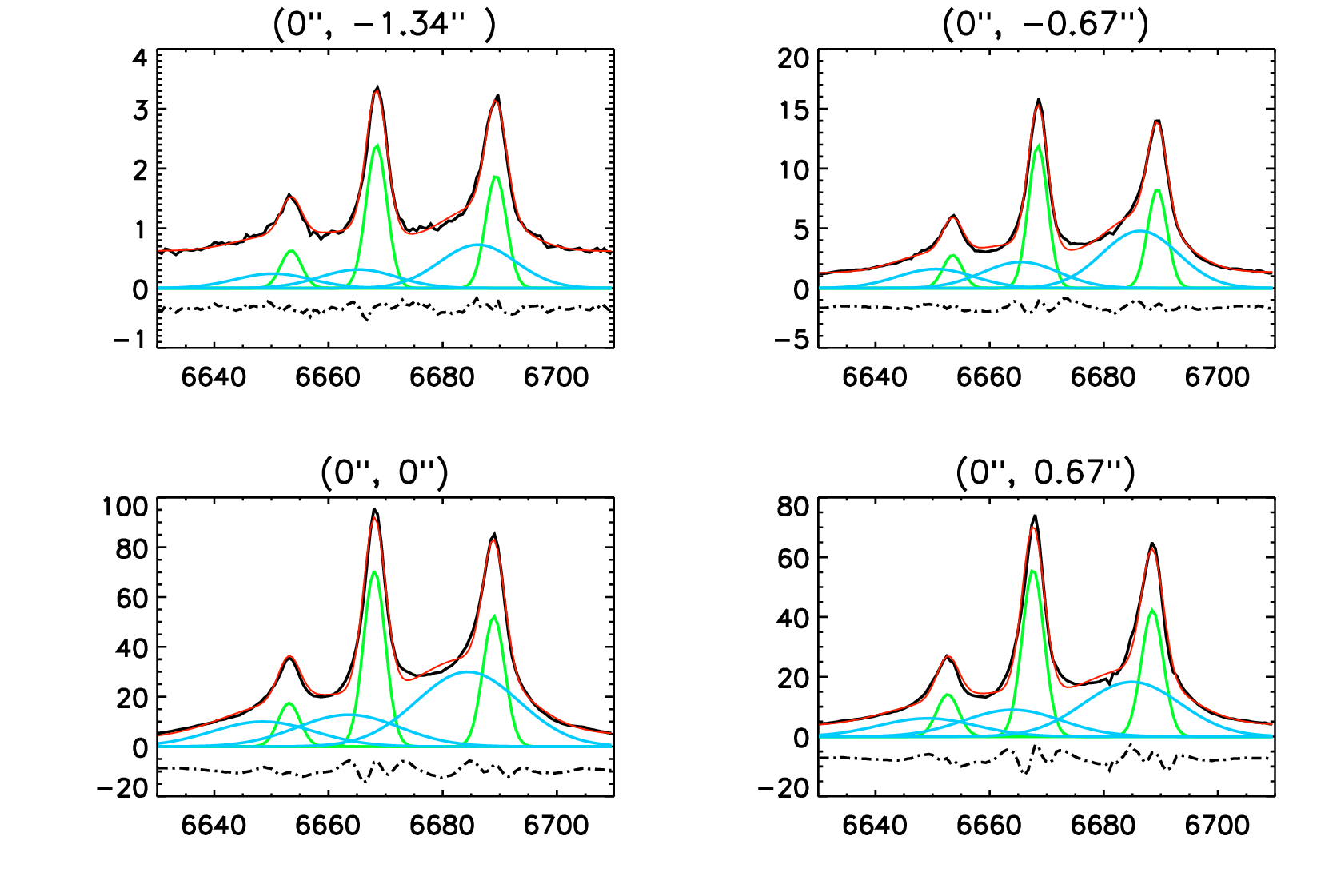}
\caption{\small{\bf Top panel:}  H$\alpha$ maps for the class 2 object {IRAS F21453-3511}. VIMOS continuum image and HST continuum images (i.e., H and I bands). {\it Middle row:} flux intensity, velocity dispersion $\sigma$ (km/s) and velocity field $v$ maps (km/s) for the main component. {\it On the bottom:} respective maps for the broad component. Note the different FoV between the main and the broad components. The latter has been zoomed in since it covers a small area. The flux intensity maps are represented in logarithmic scale and in arbitrary flux units. All the images are centered using the H$\alpha$ peak and the iso-countors of the H$\alpha$ flux are overplotted. {\bf Bottom panel:} H$\alpha$-[NII] observed spectra of {IRAS F21453-3511} for a selected inner region (indicated by the coordinates in the top label using the same reference systems as in the top figure) where main and broad component coexist. The red curve show the total H$\alpha$-[NII] components as obtained from multi-components Gaussian fits. The green and blue curves represent respectively the main and broad components.}
\label{class2}
\end{center}
\end{figure}

A small region of about 1.5 kpc x 1.3 kpc in the nuclear part shows a second component. The blue-shifted region of its velocity field shows the largest velocity dispersion (i.e., $\sigma$ $\sim$ 400 km/s, see Fig. \ref{class2}). This feature suggests that an outflow is present in the inner part of the galaxy. The amplitude of the velocity dispersion map is about $(285 \pm 20)$ km/s while the velocity field amplitude is of $(270 \pm 16)$ km/s. This component is blue-shifted by 150 km/s with respect to the main component.

The derived $v_{c}/\sigma_c$ and v$_{shear}/\Sigma$ parameters are respectively 3 and 2: in this object the random motions do not seem to dominate although in its velocity field and velocity dispersion maps there are some anomalies. With such a parameter we would classify this source as $rotation$ dominated.

\subsection{ \bf  Summary of the global kinematics properties}

From the kinematic maps and the results derived so far (see table \ref{pixel}) we can draw some general conclusions. Class 0 objects are characterized by point-antisymmetric velocity fields, showing the {\it spider diagram} typical of a ideal rotating disk; their velocity dispersion maps are centrally peaked with a radial decay at larger radii. On the other hand, class 2 objects show velocity fields which do not follow this pattern. Despite they are to first approximation point-antisymmetric, they also show clear disturbances and asymmetries. These could be explained by a bent disk and/or other tidal motions, consequence of the past merger. Their $\sigma$ maps are more asymmetric with off-nuclear regions of peculiar high velocity dispersion values (e.g., {IRAS F04315-0840}), and showing asymmetrical structures in the nuclear regions (e.g., IRAS F21453-3511).

All of our sources show signs of a broad and blue-shifted components (i.e., with velocity shift of $\Delta$v = (50 - 150) km/s with respect to the main component) in the inner part (nucleus) of the galaxy. They show mean velocity dispersions ranging between (140 - 260) km/s, which are 2 - 4 times higher than those of the main components. These components cover a small area in $disk$ objects (i.e., $\sim$ 1 kpc$^2$) while larger areas (i.e., 2 - 6 kpc$^2$) are involved for the post-coalescence $mergers$, showing more complex and distorted spectra. Although this component is more prominent for the class 2 objects, the fact that it is blueshifted for the four sources and have high associated velocities suggests that an outflow can exist in their inner regions. Furthermore, the velocity field patterns of this component (i.e., kinematic axes perpendicular to those of the main component) support the dusty outflow hypothesis.

From our kinematic results we see that the sole use of 1D parameters $v_c/\sigma_c$ and / or v$_{shear}/\Sigma$ does not seem to give us a unique classification for our sources. Indeed, according to them, our objects seem to be {\it rotation dominated,} even when the inclination correction is not included. Their v$_c/\sigma_c$ values are between those typical of local spiral galaxies (i.e., v$_c /\sigma_c \sim$ 5 - 15, \citealt{Epi10}) and those obtained for Lyman Break Analogs at z $\sim$ 0.2 (with v$_c /\sigma_c$ $\sim$ 0.4 - 1.8,  \citealt{Gon10}). The derived global velocity dispersions $\Sigma$ generally are between (40 - 60) km/s, quite comparable to those obtained in \cite{Gon10} for LBAs, with a typical median value of $\sim$ 67 km/s, but much higher than those observed in other lower luminosity local star-forming galaxies (i.e., typical $\sigma\sim$ 5 - 15 km/s, e.g., \citealt{Dib06}). For a comparison we also derived the v$_{shear}$ (not corrected for the inclination of the galaxy) as explained before. Our values ranges between 50 (i.e., IRAS F04315-0840) and 130 km/s. This parameter reveals higher degree of rotation if compared with the values obtained for LBAs (i.e., v$_{shear} <$ 70 km/s) where in many cases a significant velocity gradient has not been observed, and no actual rotation can be identified. Then, we derive v$_{shear}/\Sigma$ showing values typical of rotation dominated objects (i.e., $\sim 2 -3$, \citealt{Epi10}) with the exception of the galaxy {IRAS F04315-0840} (see Tab. \ref{pixel}).

The derived dynamical masses classify our LIRGs as moderate mass systems, characterized by a mean value of (1.8 $\pm$ 0.4) $\cdot $ 10$^{10}$ M$_\odot$. Our mass estimates agree with those obtained for LIRGs in \cite{H06} and \cite{Vais08}, which are in the range (10$^{10}$ - 10$^{11}$) M$_\odot$.

The post-coalescence mergers do not show extreme kinematic asymmetries in their maps, making them good examples to test the potential of $kinemetry$ for characterizing moderate asymmetries. 

\section {Kinemetry analysis}

The goal of this section is to investigate the potential of the $kinemetry$ method, developed by \citet{K06} (hereafter K06), when analyzing the kinematic maps of (U)LIRGs. In particular we want to investigate how powerful this methodology is for studying the kinematic asymmetries. First we will simply apply the $kinemetry$ method to the sample, drawing preliminary conclusions on the kinematics of these objects, then we will apply the same criteria as those proposed by \citet{S08} and also explore the potential of a new criterion to study kinematic asymmetries. Therefore, we expect to find out lower asymmetries for our two post-coalescence merger systems with respect to those considered by S08, which are more kinematically disturbed due to recent merger activity. In our post-coalescence systems, the inner parts are expected to be relaxed into an almost virialized disk with a large rotational component while the outer parts should still retain asymmetries associated to the merger events. Finally we analyze the resolution / redshift effects on these results.

\subsection{The method}

The $kinemetry$ method, developed by \cite{K06}, comprises a decomposition of the moment maps into Fourier components using ellipses. For clarity, we will briefly describe again the main steps presented in K06 for a better understanding of this analysis. The Fourier analysis is the most straightforward approach to characterize any periodic phenomenon: the periodicity of a  kinematic moment can easily be seen by expressing the moment in polar coordinates: K (x, y) $\rightarrow$ K (r, $\psi$). The map K(r, $\psi$) can be expanded as follows to a finite number (N+1) of harmonic terms (frequencies):

\begin{equation}
K(r, \psi) = A_{0}(r) + \sum _{n=1}^N A_{n}(r) \hspace{1mm}sin(n \cdot \psi) + B_{n}(r)\hspace{1mm} cos(n \cdot\psi) , 
\end{equation} 

where $\psi$ is the azimuthal angle in the plane of the galaxy (measured from the major axis) and $r$ is the radius of a generic the ellipse. The harmonic series can be presented in a more compact way

\begin{equation}
K(r, \psi) = A_{0}(r) + \sum _{n=1}^N k_{n}(r) \cdot cos[n(\psi-\phi_{n}(r))] , 
\label{kin}
\end{equation} 

where the amplitude and the phase coefficients ($k_{n}$, $\phi_{n}$) are easily calculated from the $A_{n}$, $B_{n}$ coefficients: $k_{n} = \sqrt{A_{n}^{2} + B_{n}^{2}}$ \hspace{1mm} and $\phi_{n} = arctan \left(\frac{A_{n}}{B_{n}}\right)$. Thus, for an \textit{ideal rotating disk} one would expect the velocity profile to be dominated by the $B_1$ term while the velocity dispersion profile dominated by the $A_0$ term. In the case of an {\it odd} moment ($\mu_{odd}$), the sampling ellipse parameters are determined by requiring that the profile along the ellipse is well described by $\mu_{odd}(\psi, r) \approx B_{1}(r) \cdot cos(\psi)$ since the velocity field peaks at the galaxy major axis ($\psi$ = 0) and goes to zero along the minor axis ($\psi$ = $\pi$/2). So, the power in the $B_{1,v}$ term therefore represents the {\it circular} velocity in each ring $r$, while power in other coefficients represent deviations from circular motion. On the other hand, the zeroth-order term, \textbf{\textit{$A_{0}$}} gives the {\it systemic velocity} of each ring. The velocity dispersion field is an {\it even} moment ($\mu_{even}$) of the velocity distribution, so that its kinematic analysis is identical to traditional surface photometry. In an ideal rotating disk the velocity dispersion has to be constant along each ring of the ellipse ($\mu_{even}(\psi, r) \approx A_0(r)$) and decreases when the semi-major axis length increases. In this context, the $A_{0}$ term represents the velocity dispersion profile. Higher order terms ($A_{n}$, $B_{n}$) will identify deviation from symmetry (\citealt{K06}).

There are two effects which limit the reliability with which coefficients in the expansion can be determined: (i) the absolute number of points sampled along the ellipse, and (ii) the regularity with which these points sample the ellipse as a function of angle, $\theta$. Ellipse parameters (i.e., position angle $\Gamma$, centre and flattening $q =b/a$) can be determined by minimizing a small number of harmonic terms; in the code some of them can be constrained (e.g., fix the center allows the position angle $\Gamma$ and flattening $q$ to freely change or remain fixed). The choice of a correct center is important in the analysis in order to avoid artificial overestimations of the asymmetries.

\subsection { {\it Kinemetry} of LIRGs sample}

In our analysis, the position of the galaxy's center is considered at the peak of the H$\alpha$  flux intensity map. In the two class 0 objects, the H$\alpha$ peak is in good positional agreement with the center of symmetry of the velocity field (i.e. kinematic center). For the class 2 objects, the kinematic center of the velocity field is not so well defined, but still the H$\alpha$ peak is in a reasonable symmetric position as well. The position angle $\Gamma$ is left free to vary. The flattening $q$ has also been left to vary, but within a physically meaningful range (i.e. 0.2-1). This allows us to consider general cases, such as tilted/wrapped disks. In Sec. 4.3.1. (i.e., Fig. \ref{Tot_plots}) we discuss further the effects of the different choices regarding these input parameters.

On figures \ref{map_rec_11255} - \ref{map_rec_21453} we present for the four galaxies the original kinematic maps (i.e., velocity field and velocity dispersion) along with their reconstructed maps (i.e., obtained using all the coefficients of the harmonic expansion up to the fifth order corrections). The residual maps (i.e., original - model) are also shown with a typical RMS of 10 km/s. Therefore, in all four cases, the fits are good and  the reconstructed maps recover the properties of the original data with great detail. W e also present the behavior of the kinematic parameters defined in the previous section (i.e., position angle, $\Gamma$, and the flattening, $q$) along with some of the $kinemetry$ coefficients (k$_1$ and k$_5$ normalized to the k$_1$).

As for the ellipse parameters, in general we find a good optimization for $\Gamma$ at the different radii for the four objects. As for $q$, there are several cases which reach the boundary of the range of acceptable values. In general this happens in the innermost regions where the rotation component, and therefore the amplitude of the sinusoidal velocity profile along the ellipse, are small. Therefore, the irregularities in the velocity field are relatively more important when compared with the (rotational) amplitude of the velocity profile. Since the problem of finding the best ellipse (i.e., best $q$ and $\Gamma$ combination) can be quite degenerate with many local minima, a low $q$ (high ellipticity) facilitates the minimization of $\chi^2$ since the latter is sensitive to a  change in $\Gamma$. On the contrary, if $q$ is high (circular ellipse), the fit ($\chi^2$) is insensitive to a change in $\Gamma$ making more difficult  the $\chi^2$ minimization process. As a result, $q$ tends to have low values in the inner regions, reaching in some cases the minimum acceptable value. In addition the $q$ parameter is less constrained at low radii as a consequence of the relatively fewer data points involved and the seeing smearing, showing larger errors\footnote{Note that $kinemetry$ provides the errors on the ellipse parameters (i.e., $q$, $\Gamma$), which are determined by the Levenberg-Marquardt least-squares minimization (MPFIT) fit with the formal 1$\sigma$ uncertainties computed from the covariance matrix. Similarly, the harmonic terms have their formal 1$\sigma$ errors estimated from the diagonal elements of the corresponding covariance matrix obtained with a linear least-square fit.}. This does not mean, however, that the harmonic expansion to describe the kinematics is not well constrained. Indeed, the reconstructed velocity maps are in excellent agreement with the real velocity field for the inner regions (i.e., low $q$ values), even for the innermost region (see figures \ref{map_rec_11255} - \ref{map_rec_21453}). For large r, the amplitude of the velocity profile along the ellipse is larger, and more data point contribute to constrain the ellipse, being less sensitive to local kinematic irregularities and, therefore,providing in general a better optimization of the parameters. At large radii we also find good agreement between the fit and the data.

It follows some comments on the $kinemetry$ results for each galaxy of the sample.

\begin{itemize}

\item {\bf IRAS F11255-4120}

The position angle $\Gamma$ is quite stable ($\sim$ 200$^\circ$ $\pm$ 15) over most the sampled radius.  In the inner part (i.e. r $< 5^{\prime\prime}$), $q$ has values of 0.2-0.4  and it has some instabilities at r $\sim$ 7$^{\prime\prime}$, likely due to the inner bar structure. However the expansion recovers very well the data in all this region. k$_1$ is mainly dominated by rotation (i.e, B$_1$) and increases radially up to a value of 140 km/s. The k$_5$/k$_1$ term, which measures the small scale kinematic asymmetries, has low values up to a maximum of 0.1, pointing some minor peculiarities for  r $<$ 7$^{\prime\prime}$, likely due to the bar structure.These asymmetries become smaller in the outer part. The error bars are not visible being smaller than the black points in the plot.

\begin{figure}[!h]
\begin{center}
\includegraphics[width=0.48\textwidth]{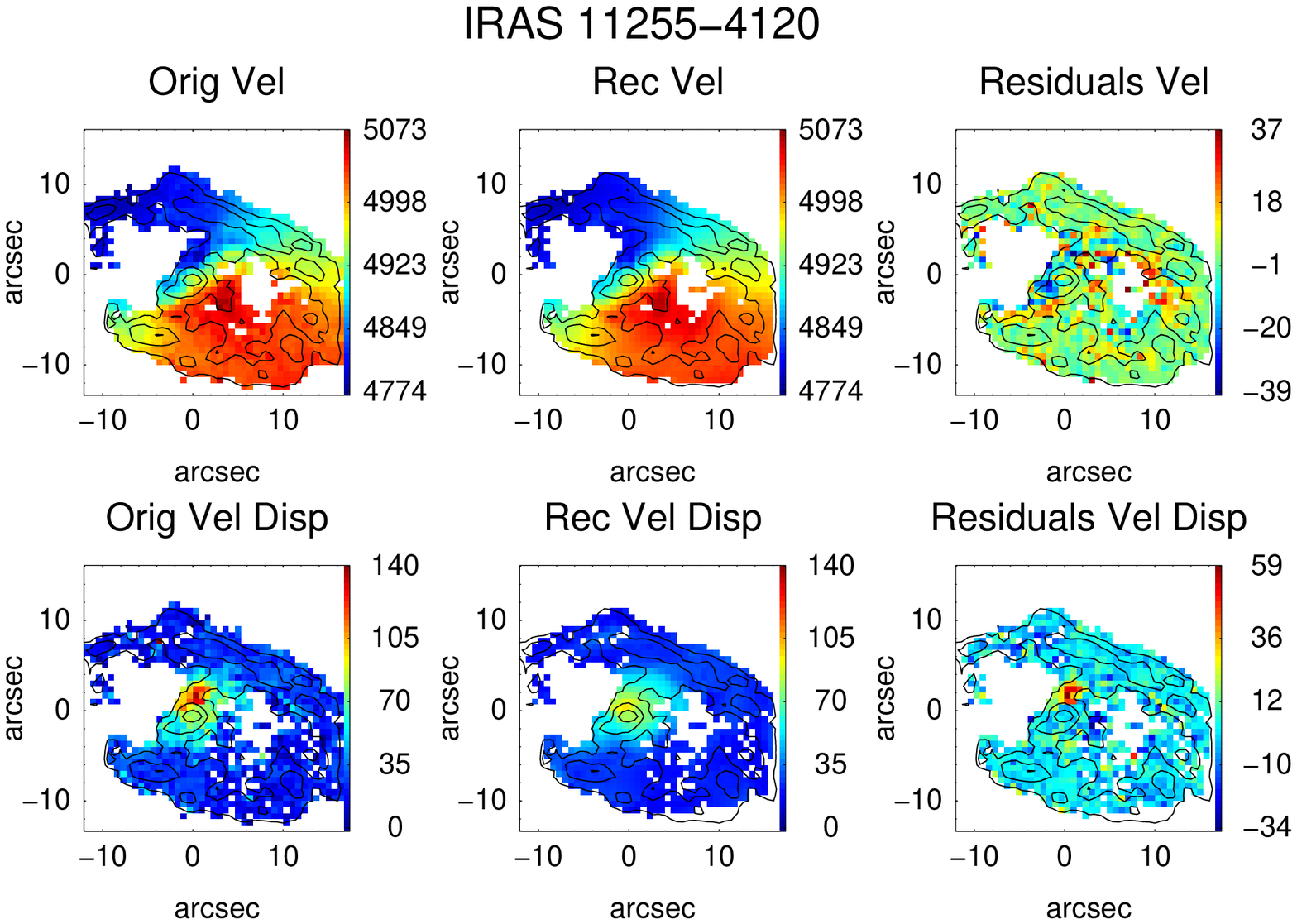}
\includegraphics[width=0.48\textwidth]{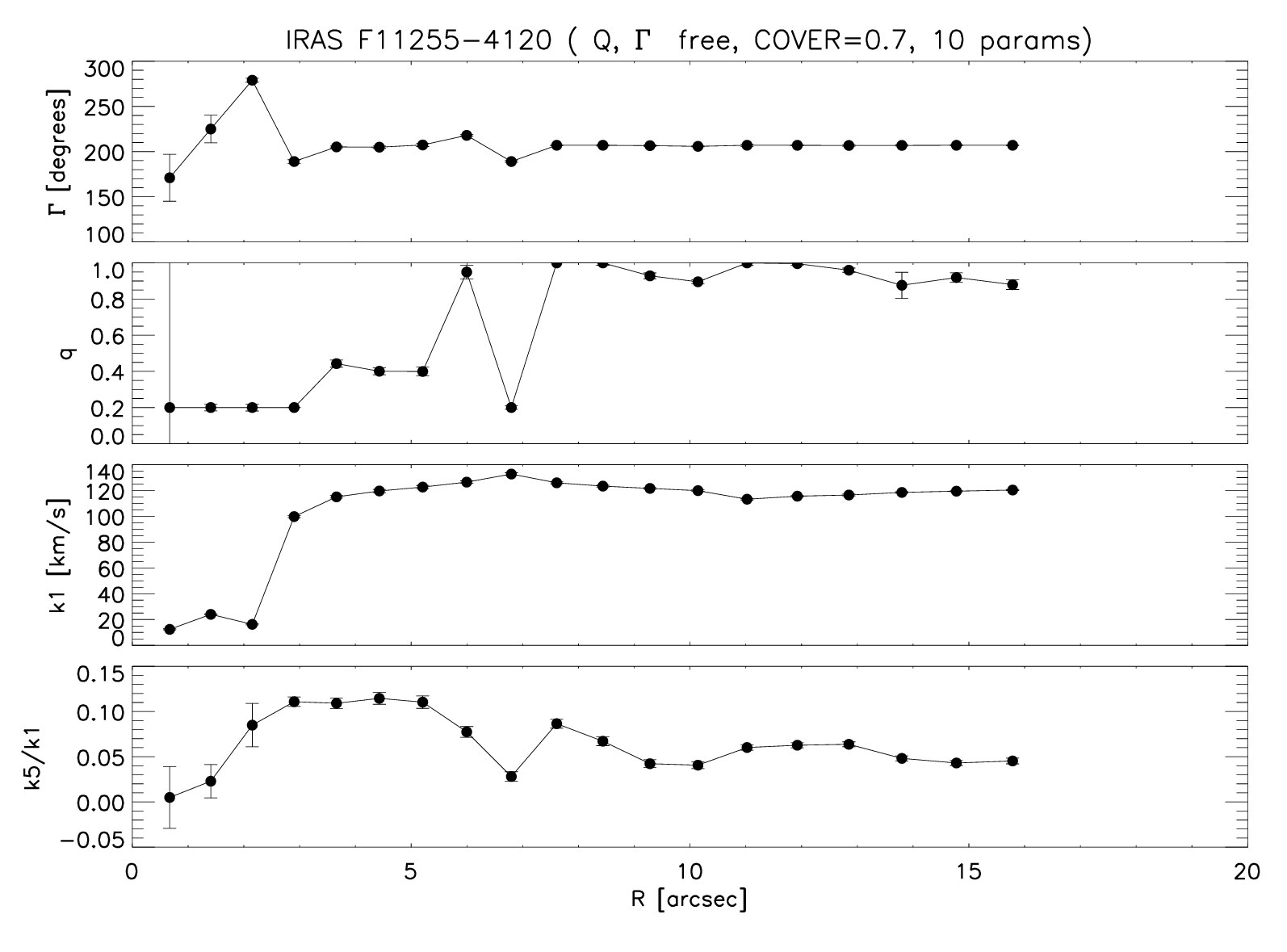}
\caption{\small {\bf Upper panel:} Maps of H$\alpha$ Gaussian fit velocities (top left), H$\alpha$ Gaussian fit dispersion (bottom left) and their respective reconstructed (middle) and residual (data-model) maps (on the right) for {IRAS F11255-4120}. {\bf Lower panel:} Radial profiles of the kinematic properties, obtained using $kinemetry$ program. The position angle $\Gamma$ and the flattening $q$ of the best fitting ellipses as well as the first and the fifth order Fourier terms (respectively, k$_1$ and k$_5$) are plotted as a function of the radius.}
\label{map_rec_11255}
\end{center}
\end{figure}

\vspace{5mm}
\item{\bf IRAS F10567-4310}

The position angle $\Gamma$ is quite stable over the sampled radii, with values between 210 and 250 degrees. The $q$ parameter reaches the value of 0.2 in the inner region (i.e., r $< 4^{\prime\prime}$), where the reconstructed map shows a very good agreement with the data. k$_1$ is mainly dominated by rotation (i.e, B$_1$) and increases radially up to a value of 140 km/s, revealing a possible minor anomaly at r $\sim$ 5$^{\prime\prime}$. The k$_5$/k$_1$ term has very low values, with a maximum of 0.05 over the sampled radii, meaning that this object is close to an ideal rotating disk structure. 

\begin{figure}[!h]
\begin{center}
\includegraphics[width=0.48\textwidth]{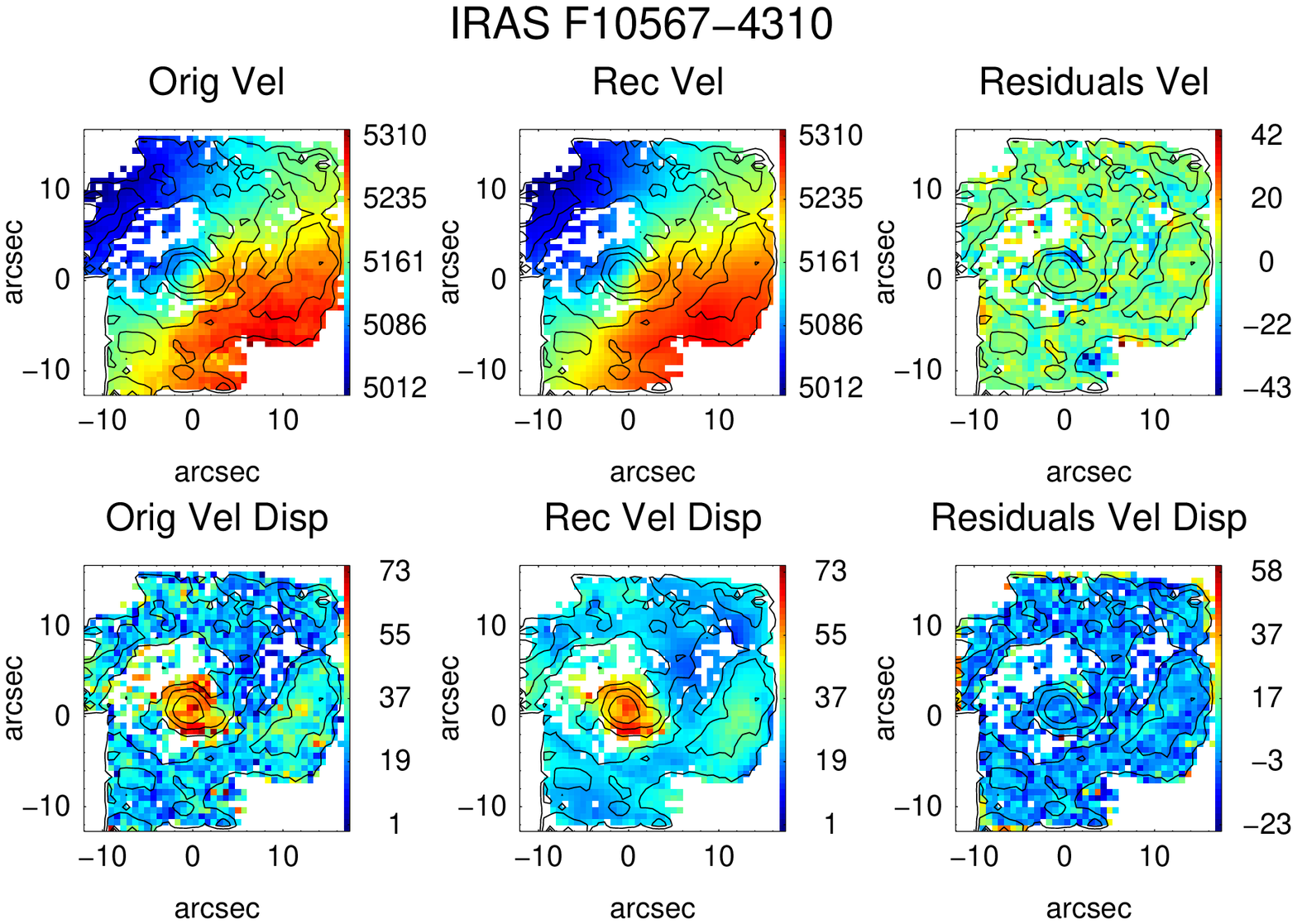}
\includegraphics[width=0.48\textwidth]{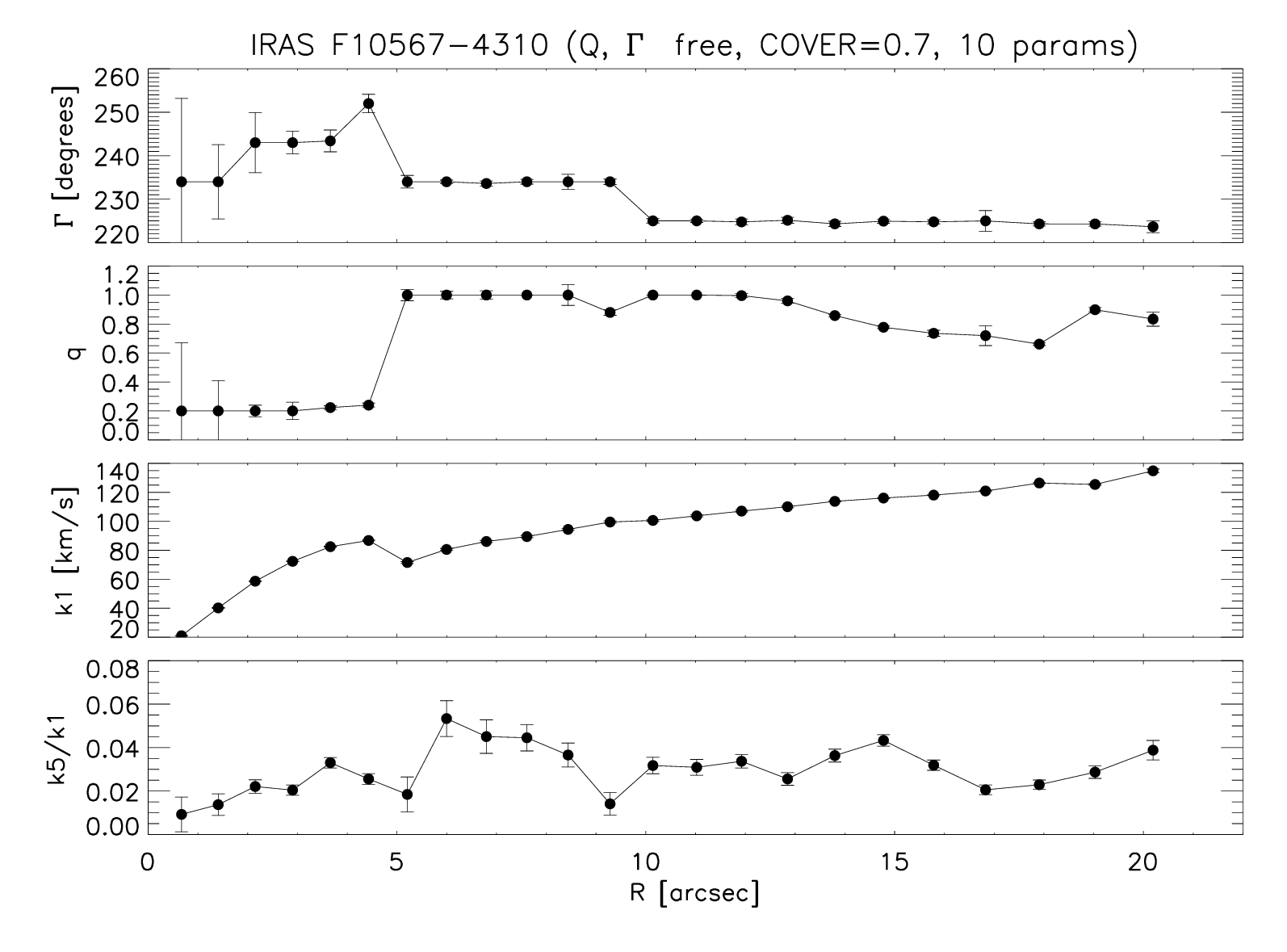}
\caption{\small {\bf Upper panel:} Maps of H$\alpha$ Gaussian fit velocities (top left), H$\alpha$ Gaussian fit dispersion (bottom left) and their respective reconstructed (middle) and residual (data-model) maps (on the right) for {IRAS F10567-4310}. {\bf Lower panel:} Radial profiles of the kinematic properties, obtained using $kinemetry$ program. The position angle $\Gamma$ and the flattening $q$ of the best fitting ellipses as well as the first and the fifth order Fourier terms (respectively, k$_1$ and k$_5$) are plotted as a function of the radius.}
\label{map_rec_10567}
\end{center}
\end{figure}

\vspace{5mm}

\item{\bf IRAS F04315-0840}

The position angle $\Gamma$ typically spans from 350$^\circ$ up to 380$^\circ$\footnote{The position angle is measured from the North (0$^\circ$=360$^\circ$) anti-clockwise. In order to make the two values more easily comparable we choose to add 360$^\circ$ in one case.}, except for (r $\sim$ 3$^{\prime\prime}$) which drops to around 320$^\circ$ as a consequence of one of the two distinct redshifted peaks present in the velocity field map. For most of the radii $q$ is close to 1(circle-ellipses). k$_1$ increases radially up to a value of 120 km/s ($r< 5^{\prime\prime}$) and then decreases to reach a value of 100 km/s at r $\sim$ 15$^{\prime\prime}$. The k$_5$/k$_1$ term is between 0 and 0.3, with an constant increasing behaviour. This illustrates that departures from rotation are mainly found in the outer parts.

\begin{figure}[!h]
\begin{center}
\includegraphics[width=0.48\textwidth]{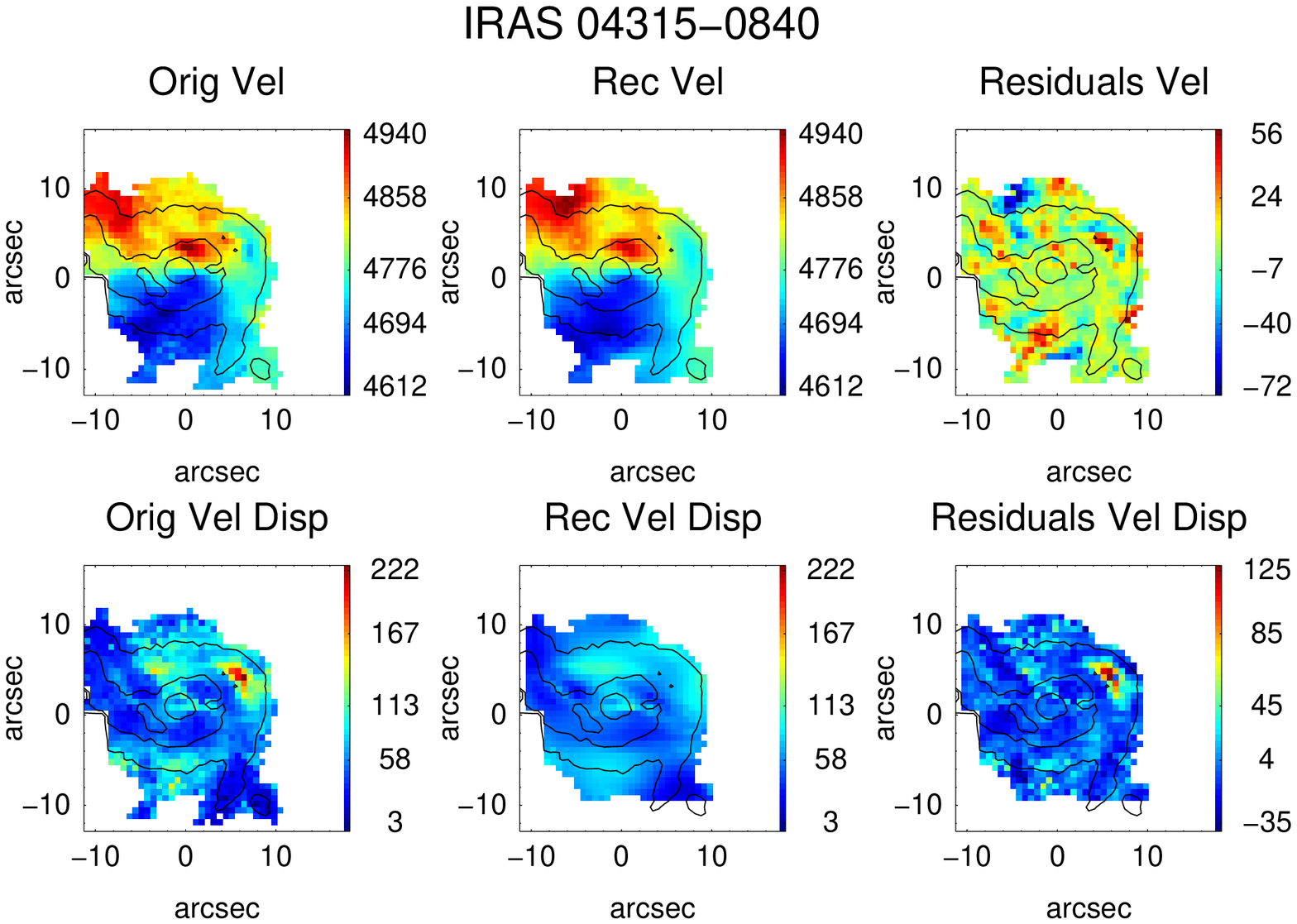}
\includegraphics[width=0.48\textwidth]{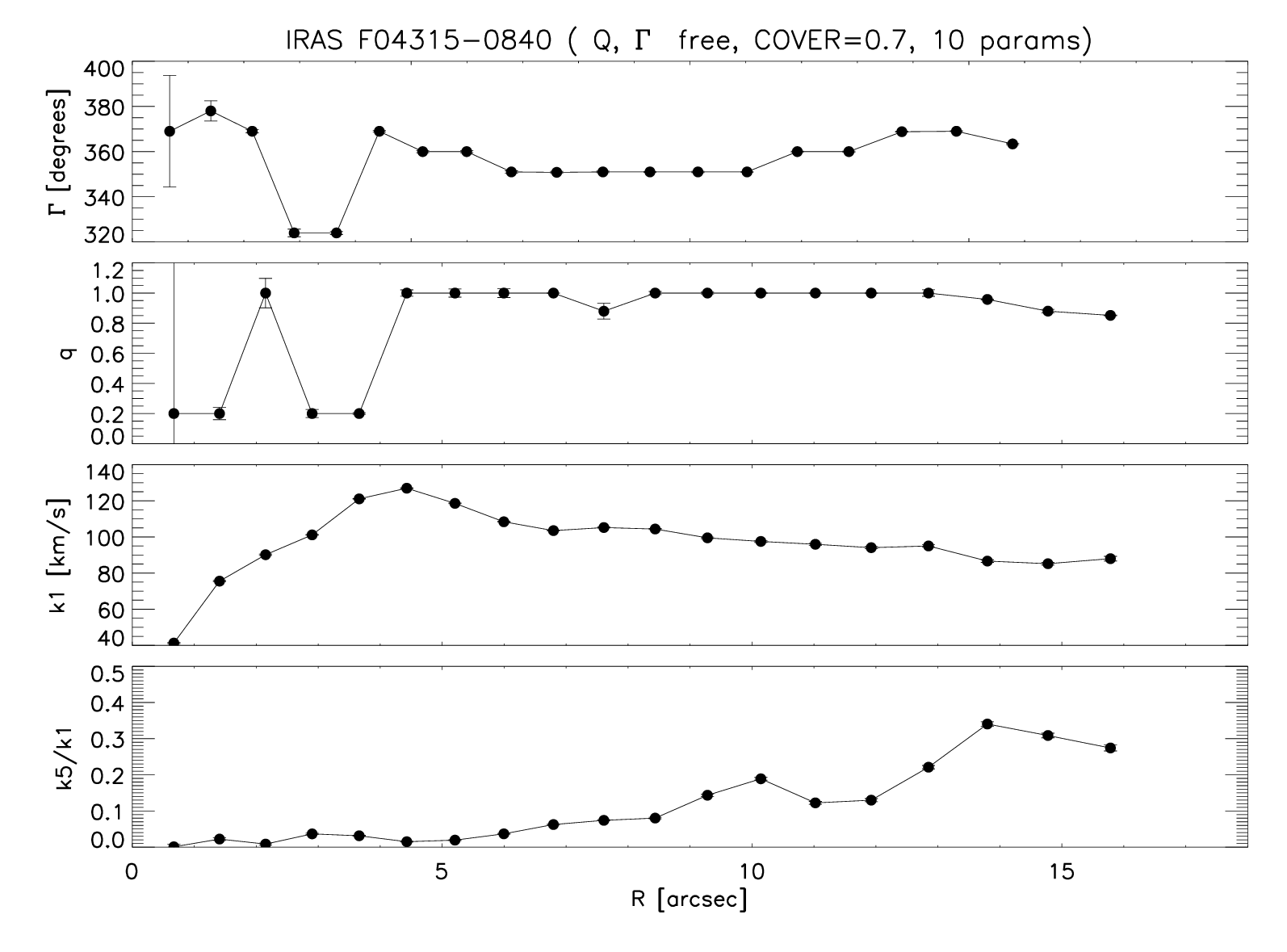}
\caption{\small {\bf Upper panel:} Maps of H$\alpha$ Gaussian fit velocities (top left), H$\alpha$ Gaussian fit dispersion (bottom left) and their respective reconstructed (middle) and residual (data-model) maps (on the right) for {IRAS F04315-0840}. {\bf Lower panel:} Radial profiles of the kinematic properties, obtained using $kinemetry$ program. The position angle $\Gamma$ and the flattening $q$ of the best fitting ellipses as well as the first and the fifth order Fourier terms (respectively, k$_1$ and k$_5$) are plotted as a function of the radius.}
\label{map_rec_04315}
\end{center}
\end{figure}

\vspace{5mm}

\item{\bf IRAS F21453-3511}

The position angle $\Gamma$ almost keeps constant to $\sim$ 240$^\circ$ for r $>$~6$^{\prime\prime}$ while it shows lower and more unstable values for r $<$~6$^{\prime\prime}$. The $q$ reaches 0.2 in the region inside a radius of r $<$~6$^{\prime\prime}$, mainly motivated by the relatively small rotation component in this region. However the agreement between the kinematic fit and data is very good for these radii. At r =7$^{\prime\prime}$ $q$ reaches a maximum ($\sim$ 1) to decrease constantly to a value of 0.4. k$_1$ does not reveal any strong rotational component ranging between 10 up to 40 km/s. It increases in the inner region for $r<4^{\prime\prime}$ and then remains constant. The k$_5$/k$_1$ term ranges between 0 and 0.2, showing a quite stable trend. For radii larger than 10$^{\prime\prime}$ it shows an opposite behavior with respect to the k$_1$ term. In this object the deviations from pure rotation are not as high as those of IRAS F04315-0840, but they are higher than those for the class 0 objects.

\begin{figure}[!h]
\begin{center}
\includegraphics[width=0.48\textwidth]{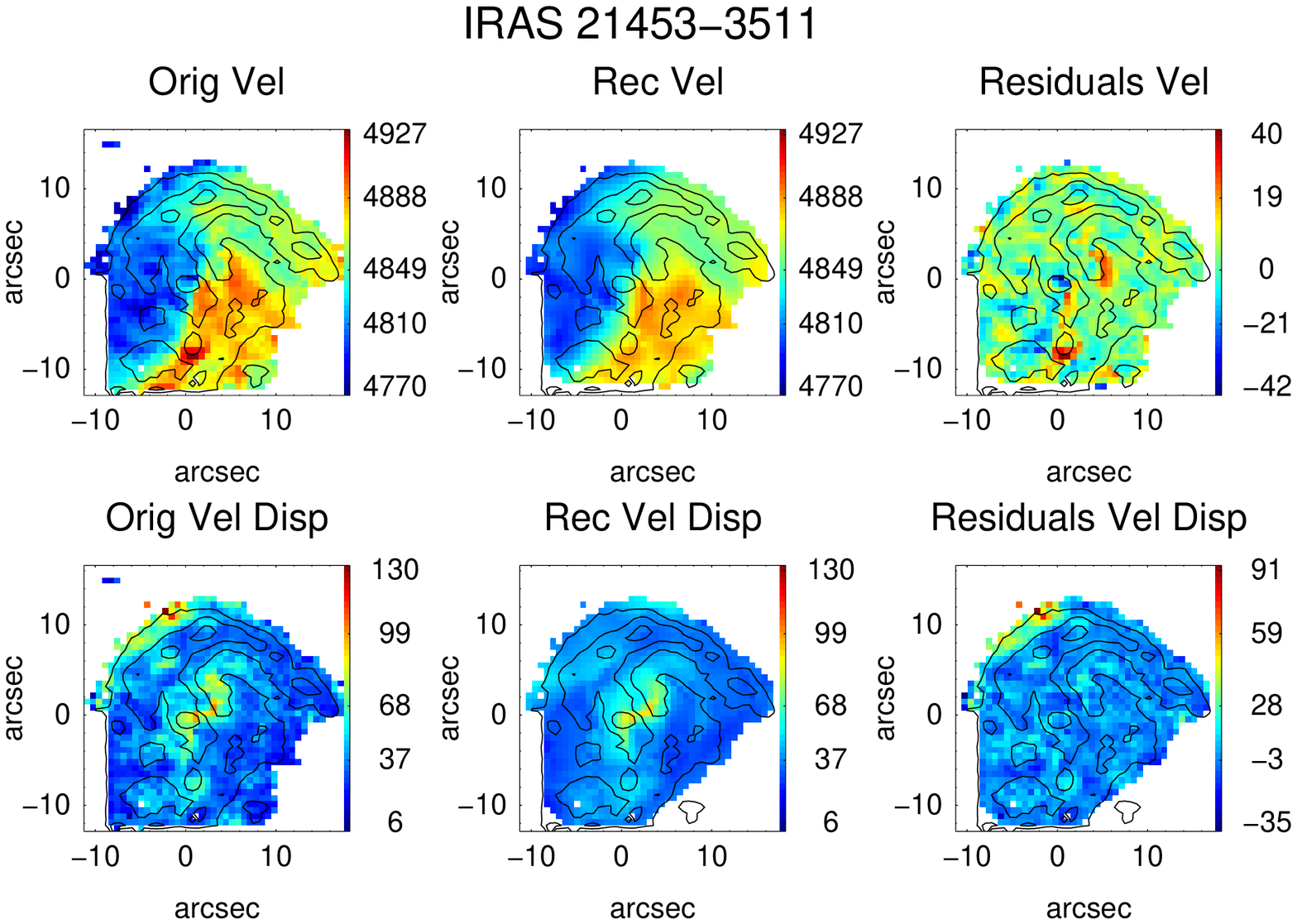}
\includegraphics[width=0.48\textwidth]{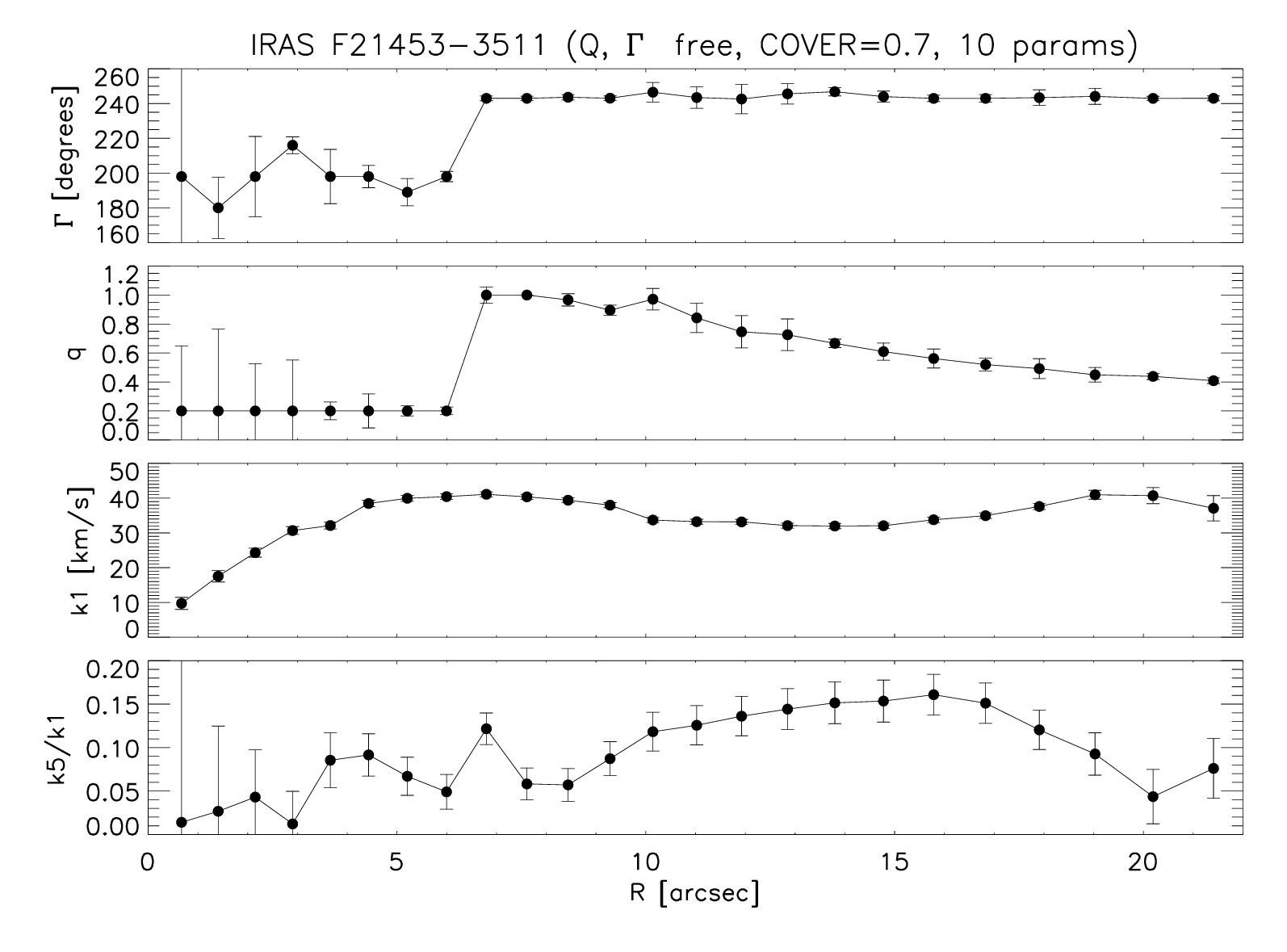}
\caption{\small {\bf Upper panel:} Maps of H$\alpha$ Gaussian fit velocities (top left), H$\alpha$ Gaussian fit dispersion (bottom left) and their respective reconstructed (middle) and residual (data-model) maps (on the right) for {IRAS  F21453-3511}. {\bf Lower panel:} Radial profiles of the kinematic properties, obtained using $kinemetry$ program. The position angle $\Gamma$ and the flattening $q$ of the best fitting ellipses as well as the first and the fifth order Fourier terms (respectively, k$_1$ and k$_5$) are plotted as a function of the radius.}
\label{map_rec_21453}
\end{center}
\end{figure}

\end{itemize}

In general we have found that for class 0 sources the higher order deviations (from pure rotation) are small (i.e., k$_5$/k$_1<$ 0.1) while for class 2 objects the deviations are higher, mainly in the outer regions (k$_5$/k$_1$ $\leq$ 0.4). For all of them the rotation curve (i.e., k$_1$ parameter) seem to characterize these objects as rotating. As discussed before, these galaxies do not show extreme asymmetries in their kinematic maps and this is confirmed by the $kinemetry$ results obtained so far.

In the following sections we will study several kinematic criteria with the aim of better classifying these systems.

\subsection {\textbf{Sample of local  LIRG systems in} the [$\sigma_a$ - v$_a$] plane }

In order to reveal the presence of rotational/non-rotational motions within the dynamics of the gas in each galaxy we will first consider the same criteria as the one proposed by S08. It is worth mentioning that they compare two main classes of systems: those which have suffered a recent major merger event (i.e., \textit{mergers}) and those without sings of interacting or merger activity (i.e., \textit{disks}). For further details see S08. They define the asymmetries in the velocity and velocity dispersion fields as:

\begin{equation}
\hspace{1cm} v_{asym} =   \left\langle \frac{ k_{avg, v} }  {B_{1, v} }  \right\rangle_r   \hspace{1cm} {\sigma_{asym} =\left\langle \frac{k_{avg, \sigma}}{B_{1, v}}\right\rangle _r}, 
\end{equation}

\vspace{5mm}

where $k_{avg, v} =(k_{2, v} + k_{3, v} + k_{4, v} + k_{5, v})/4$  and $k_{avg, \sigma} =(k_{1, \sigma} + k_{2, \sigma} + k_{3, \sigma} + k_{4, \sigma} + k_{5, \sigma})/5$. For an ideal rotating disk, we expect the velocity profile to be perfectly antisymmetric where the $B_1$ term would dominate the Fourier expansion, while the velocity dispersion map is expected to be perfectly symmetric and therefore all terms except $A_0$ would vanish.

\begin{figure}
\begin{center}
\includegraphics[width=0.48\textwidth]{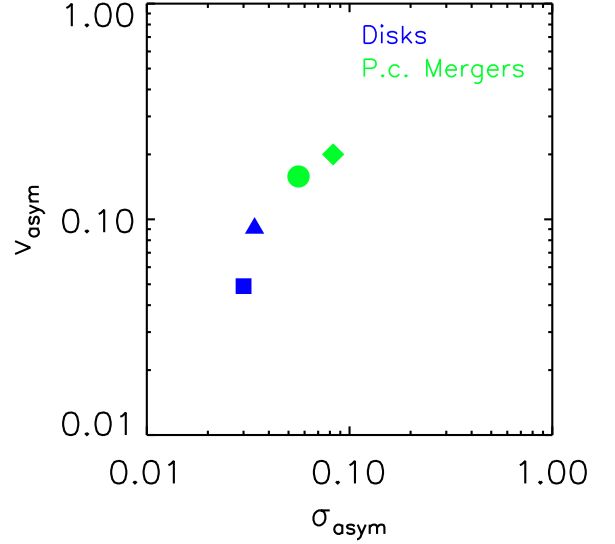}
\caption{\small Asymmetry measure of the velocity v$_{asym}$ and velocity dispersion $\sigma_{asym}$ fields for the four galaxies analyzed here. Symbol types distinguish among different systems: the square represents {IRAS F10567-4310}, the triangle is IRAS F11255-4120, the circle is {IRAS F04315-0840} and diamond represents {IRAS F21453-3511}. Blue and green symbols stand for different morphological types, respectively class 0 and class 2 objects.}	
\label{Shap_10}
\end{center}
\end{figure}

In Fig. \ref{Shap_10} we show the results for our galaxies in the [$\sigma_a - v_a$] plane. As expected, class 0 objects have lower values of the asymmetries than class 2 objects. Indeed, looking at their velocity and velocity dispersion maps, the class 0 maps (Figs. \ref{map_rec_11255} - \ref{map_rec_10567}) resemble to those of an ideal rotating disk (i.e., `spider diagram' structure for the velocity field and centrally peaked velocity dispersion map) while class 2 objects present more distorted velocity fields and irregular dispersion maps (Figs. \ref{map_rec_04315} - \ref{map_rec_21453}). Therefore, taking into account that the objects in the present sample were classified as class 0 or class 2 on the basis of pure morphological arguments, we can conclude that their morphological and kinematics classification in the [$\sigma_a$ - v$_a$] plane are consistent.

In order to analyze the robustness of the results in Fig. \ref{Shap_10}, we will analyze in the following sections their dependence on the input parameters considered as well as in the error in the radial velocities and velocity dispersion measurements. 
Note that $kinemetry$ uses the full 2D kinematic information of the velocity field and velocity dispersion map, allowing us a good characterization of the asymmetries, something which is more efficient than using 1D parameters such as v$_c/\sigma_c$ and v$_{shear}/\Sigma$ as seen in section 3.

\subsubsection{Dependence on the input parameters}
\label{constrains}

As pointed out by K06, a good choice of input parameters is important in order to avoid an artificial overestimation of the asymmetries. To perform a $kinemetry$ analysis we require to set the dynamical center, as well as to specify different levels of constrains of the input parameters. For instance, the kinematic position angle $\Gamma$ and the flattening $q$ of the ellipses can be fixed or left free to vary for the fitting at different radii. The COVER parameter, which controls the radius at which the process stops by setting the fraction of the ellipse that has to be covered by data\footnote{For instance, COVER = 0.7 means that if less than 70\% of the points along an ellipse are not  covered by data the program stops. This value makes sure that kinematic coefficients are robust. Sometimes it is necessary to relax this condition especially when reconstructing maps. See further details in \cite{K06}.}, has to be assigned too.

\begin{figure}
\begin{center}
\includegraphics[width=0.48\textwidth]{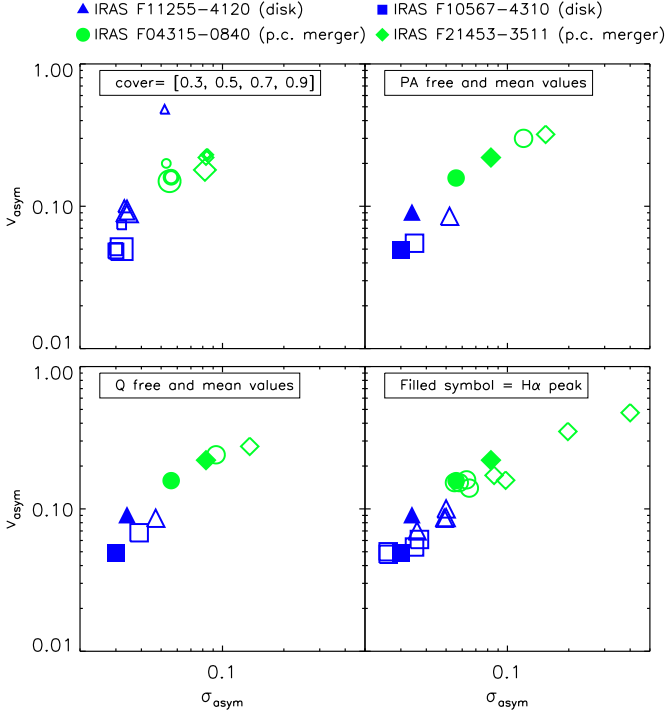}
\caption{\small Results for the asymmetry measures when different sets of input parameters are considered with 10 harmonic terms analysis. Symbol types and colors distinguish among the different systems as explain in the legend. In each panel one parameter at a time is changed: respectively, COVER, $\Gamma$ (or PA), $q$ and the CENTER of the ellipses. {\it On the top-left:} Different results for the COVER parameter (i.e., 0.3, 0.5, 0.7, 0.9) where the size of the symbol is proportional to the COVER value (the biggest symbol corresponds to the highest cover value and vice versa). {\it On the top-right:} The results obtained considering a {\it position angle} $\Gamma$ free to vary (filled symbols) or fixed to its mean value (empty symbols) are shown. {\it On the bottom-left:} The results shown are obtained when considering $q$ free to vary and constant with the radius to its mean value. {\it On the bottom-right:} Results achieved choosing 5 different centers for each galaxy:  filled symbols represent results obtained from the `standard' analysis (H$\alpha$ flux peak), empty symbols the results from shifting their center of 1 pixel (horizontally {\it and} vertically, with respect to the H$\alpha$ peak pixel corresponding to a shift of 0.95$^{\prime\prime}$). }  
\label{Tot_plots}
\end{center}
\end{figure}

 In Fig. \ref{Tot_plots} we present the results in the [$\sigma_a$ - v$_a$] plane for our galaxies for different sets of input parameters. In each panel one parameter at a time is changed: respectively, the COVER (top-left), the position angle $\Gamma$ (or PA, top-right), the flattening $q$ (bottom-left) and the CENTER (bottom-right) of the ellipses. We can find that the results are stable to a reasonable choice of values. Indeed, looking at the COVER panel, the four galaxies give similar results up to cover=0.5; {IRAS F11255-4120} deviates significantly for COVER=0.3. {IRAS F21453-3511} seems to be sensitive to the galaxy center but it maintains within the region of high-asymmetries in all the (extreme) cases considered. As shown, the choice of free or fixed position angle $\Gamma$ or flattening $q$ do not affect so much the final results.  In particular the computed asymmetries are quite insensitive to the choice for $q$, especially for the class 0 galaxies.  In general, the results obtained with a fixed $\Gamma$ / $q$ are somewhat  higher due to the fact that there is less degree of freedom such that the asymmetries cannot be well accounted by the fitting.  Obviously the proper choice of free / fixed parameters depends on the level of S/N (i.e., if the deviations are associated to a true feature or noise).
 
According to these results we have selected the following set of input parameters for the remainder analysis (i.e., COVER = 0.7, $\Gamma$ is completely free to vary and $q$ free to vary in the range [$0.2-1$], center = H$\alpha$ flux peak).

\subsubsection{Monte Carlo simulations}

In order to analyze the dependence of the $kinemetry$ results to the uncertainties in the radial velocities and velocity dispersion values, we measure the probability distribution function (PDFs) of the asymmetries in these systems using Monte Carlo (MC) simulations, as done in S08, since the $kinemetry$ method does not lend itself to a straightforward error propagation.

\begin{figure}[h]
\begin{center}
\includegraphics[width=0.45\textwidth]{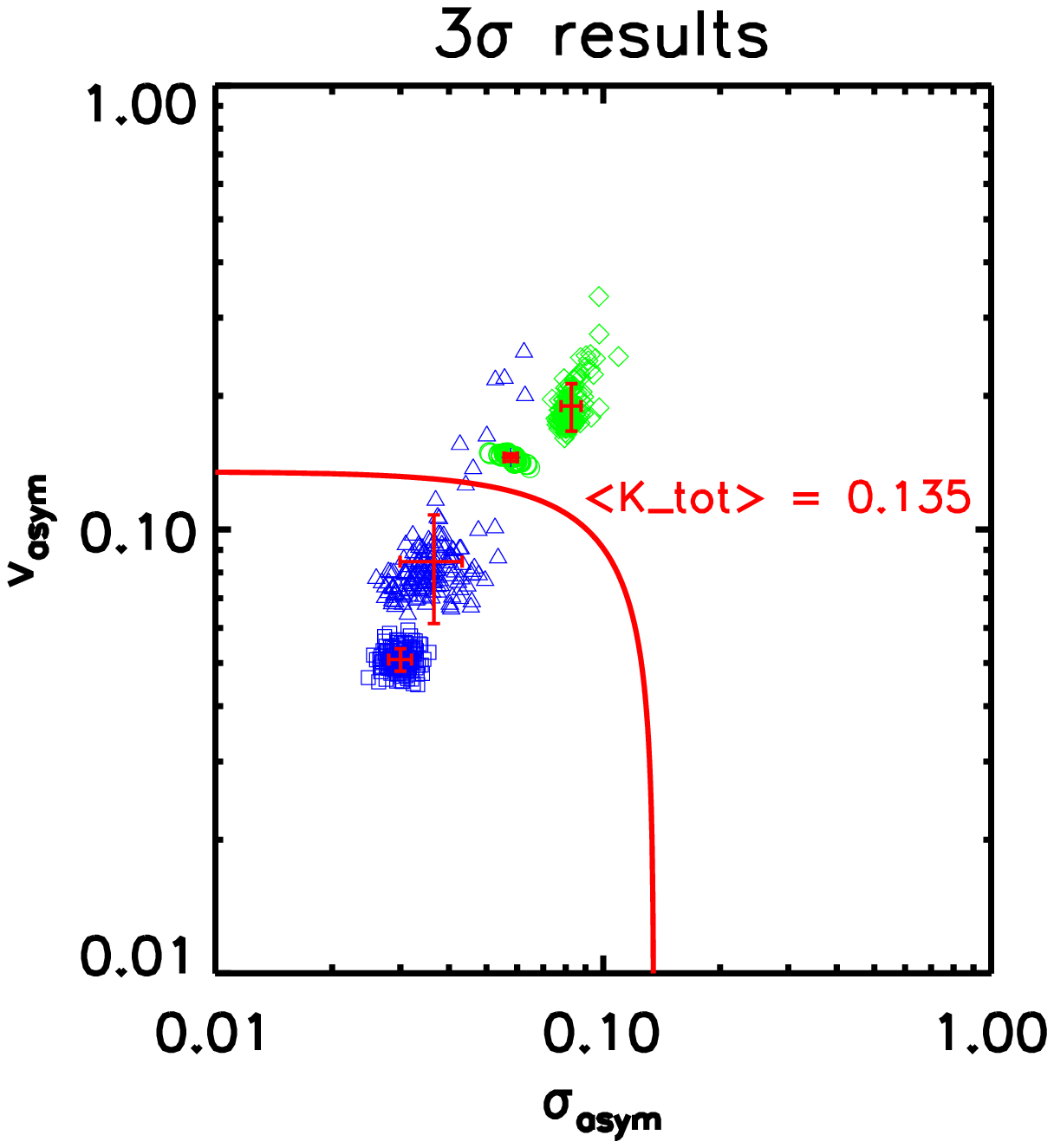}
\includegraphics[width=0.4\textwidth, angle=90]{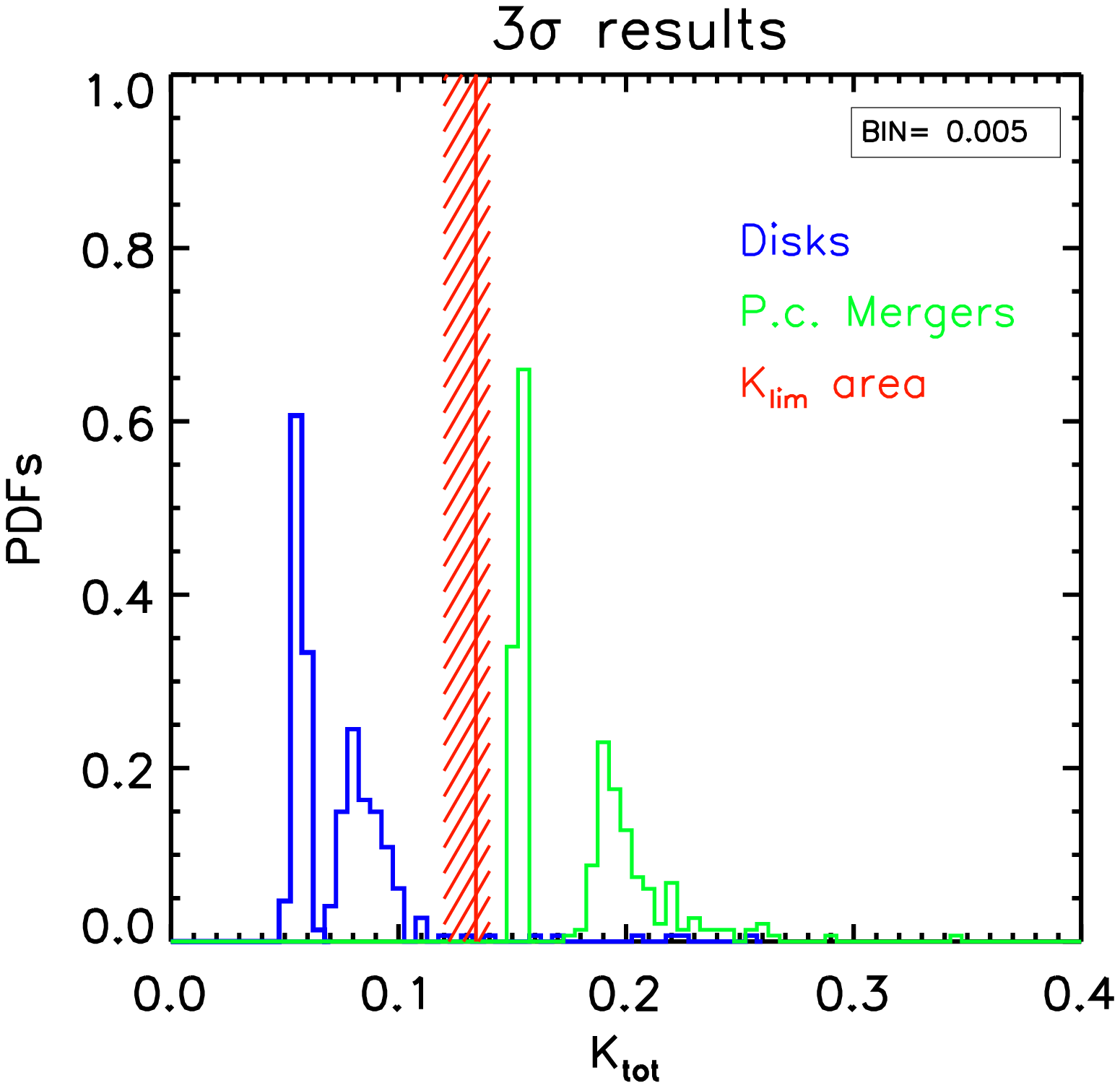} 
\caption{\small {\bf On the top:} Asymmetry measures of the velocity v$_{asym}$ and velocity dispersion $\sigma_{asym}$ fields as derived from the Monte Carlo realizations for the four objects. For each source 150 MC simulations are run but here only the $\pm$3$\sigma$ results are shown. The solid red line indicates the division between class 0 (disks) and class 2 (post-coalescence mergers) at $<K_{tot}>$ = 0.135. {\bf On the bottom:} The probability distribution function (PDFs) as derived from MC realizations. The empirical delineation where $K_{lim}=0.135$ can cleanly separate the two classes (i.e., $K_{lim}$ area). The symbols are the same as the ones used before.}
\label{MC_1}
\end{center}
\end{figure}

Therefore, for each template, we create 150 different realizations of the moment maps (i.e., velocity field and velocity dispersion) based on their corresponding error maps. These error maps correspond to the measurement errors of the velocity moments, as derived when fitting the kinematics from the data cube along with wavelength calibration errors, as explained in section 2.4.  For each moment map we perturb the observed data points by randomizing them, using Gaussian noise parameterized by the measured 1$\sigma$ errors. The new maps created are then used to rerun \textit{kinemetry} and apply the analysis described before. In figure \ref{MC_1} the results are shown. They show that the results in the [$\sigma_a$ - v$_a$] plane are stable to the velocity errors, with relatively well defined regions for disks and merger galaxies.
	
\subsection{Total kinematic asymmetry of disk, on-going and post-coalescence merger systems }

As done in S08 we compute the total kinematic asymmetry K$_{tot}$ border in order to separate class 0 (disks) and class 2 (post-coalescence merger) objects. For the four objects we obtained a mean value of $<K_{tot}> = 0.135$: this is a significant lower value than that obtained in S08 (i.e., $<K_{tot}> = 0.5$). The difference could arise from the fact that our \textit{mergers} are in a post-coalescence phase showing relaxation in the innermost regions with a large rotation component, while in S08 they are mostly pre-coalescence merger systems dominated by dispersion. Also, the limit defined in S08 comes out after excluding {\it IRAS 12112+0305}, a pre-coalescence merger pair (i.e., \citealt{GM09}) that looks like a \textit{disk} at high redshift. If the classification of this object is considered, the total kinematic asymmetry border derived for the whole S08 sample would have been considerably lower (i.e., $<K_{tot}>  \sim$ 0.3, see Figure 5 in S08) and therefore the discrepancy with our finding reduced.

In any case, the fact that the border $disk/merger$ for the S08 sample is so dependent on the classification of a single object (i.e., IRAS 12112+0305) illustrates its relatively large associated uncertainties. A reduction of the total asymmetry border K$_{tot}$, as our results (and S08, after reclassifying {\it IRAS 12112+0305}) suggests, would imply that the relative frequency of $disks$ and $mergers$ in a given sample changes, increasing the fraction of $mergers$. Changing the relative frequencies of $disk/merger$ have obvious implications when interpreting the data in terms of the different evolutionary scenario mentioned in the introduction. On the other hand our sample is admittedly too small and formed by objects with relatively homogeneous properties (i.e., LIRGs classified as disks and post-coalescence mergers) for a robust determination of general use. Therefore, further efforts to constrain its value as well as to understand how it depends on several instrumental and observational factors are required.

\subsection {A new criterion to distinguish kinematic asymmetries between post-coalescence mergers from disks}

In order to better assess the presence of asymmetries in the kinematic maps we will explore a new kinematic criterion. 

As described in \cite{kron07}, when considering recent (i.e., $\leq$ 100 Myr after the first encounter) or ongoing major mergers of equal mass galaxies (i.e., Milky Way type), the inner regions\footnote{If a galaxy covers a FoV of 30 kpc $\times$ 30 kpc, the regions more affected by chaotic motions are those at galactocentric distances smaller than 10 kpc.} of the galaxy are usually affected by more chaotic motions, as revealed by a quite irregular rotational curve and higher order deviations (i.e., k$_5$/k$_1$) at small radii. As the major merger evolves, the inner regions rapidly relax into a rotating disk, while the outer parts are still out of equilibrium. This implies that the velocity field in a post-coalescence system could be dominated by rotation in the inner regions with large kinematic asymmetries in the outer parts, as it is actually observed in our systems. Provided that the outer regions retain better the memory of a merger event, we propose a criterion which enhances the relative importance of the asymmetries  at larger radii. 

Indeed, instead of simply averaging the asymmetries over all radii (as in S08), these are weighted according to the number of data points used in their determination. Since the number of data points of the outer ellipses is larger than for the inner ones, the asymmetries found in the outer ellipses contribute more significantly to the average for obtaining v$_{asym}$ and $\sigma_{asym}$. As the number of data points is in first approximation proportional to the circumference of the ellipse, for practical reasons we used this to weight the asymmetries found for the different ellipses. 

The circumferences of the ellipses are computed using the truncated `infinite sum' formula, that is a function of the ellipticity (i.e., $e = \sqrt{1-q^2}$) and the semi major axis of the ellipse (r):
 
\begin{equation}
\hspace{5mm}C(e, r) \approx 2 \pi r \left[1- \left(\frac{1}{2}\right)^2 e^2 - \left(\frac{1\cdot 3}{2 \cdot 4}\right)^2 \cdot \frac{e^4}{3} \right]
\label{peri}
\end{equation} 

 The final formula to compute the weighted velocity and velocity dispersion asymmetries are respectively:

\begin{equation}
\hspace{5mm}v_{asym} = \sum_{n=1}^N  \left (\frac{ k_{avg, n}^v}{B_{1, n}^v } \cdot C_n \right )\cdot \frac{1}{\sum_{n=1}^N C_n} 
\label{v_weighted}
\end{equation}

\begin{equation}
\hspace{5mm}\sigma_{asym} = \sum_{n=1}^N  \left ( \frac{ k_{avg, n}^\sigma}{B_{1, n}^v } \cdot C_n \right )\cdot \frac{1}{\sum_{n=1}^N C_n} 
\label{s_weighted}
\end{equation} 

\begin{figure}[h]
\begin{center}
\includegraphics[width=0.45\textwidth]{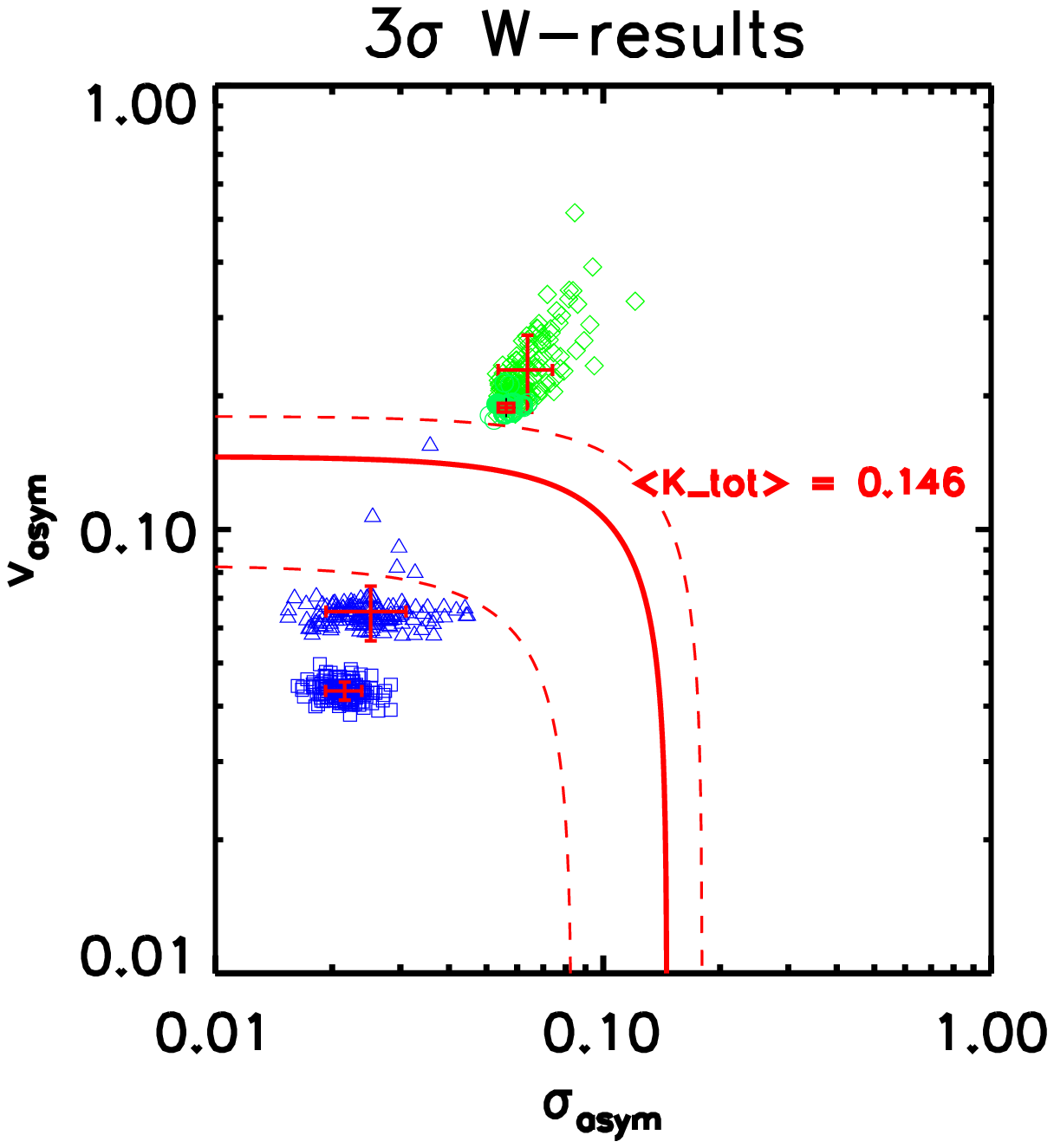}
\includegraphics[width=0.4\textwidth, angle=90]{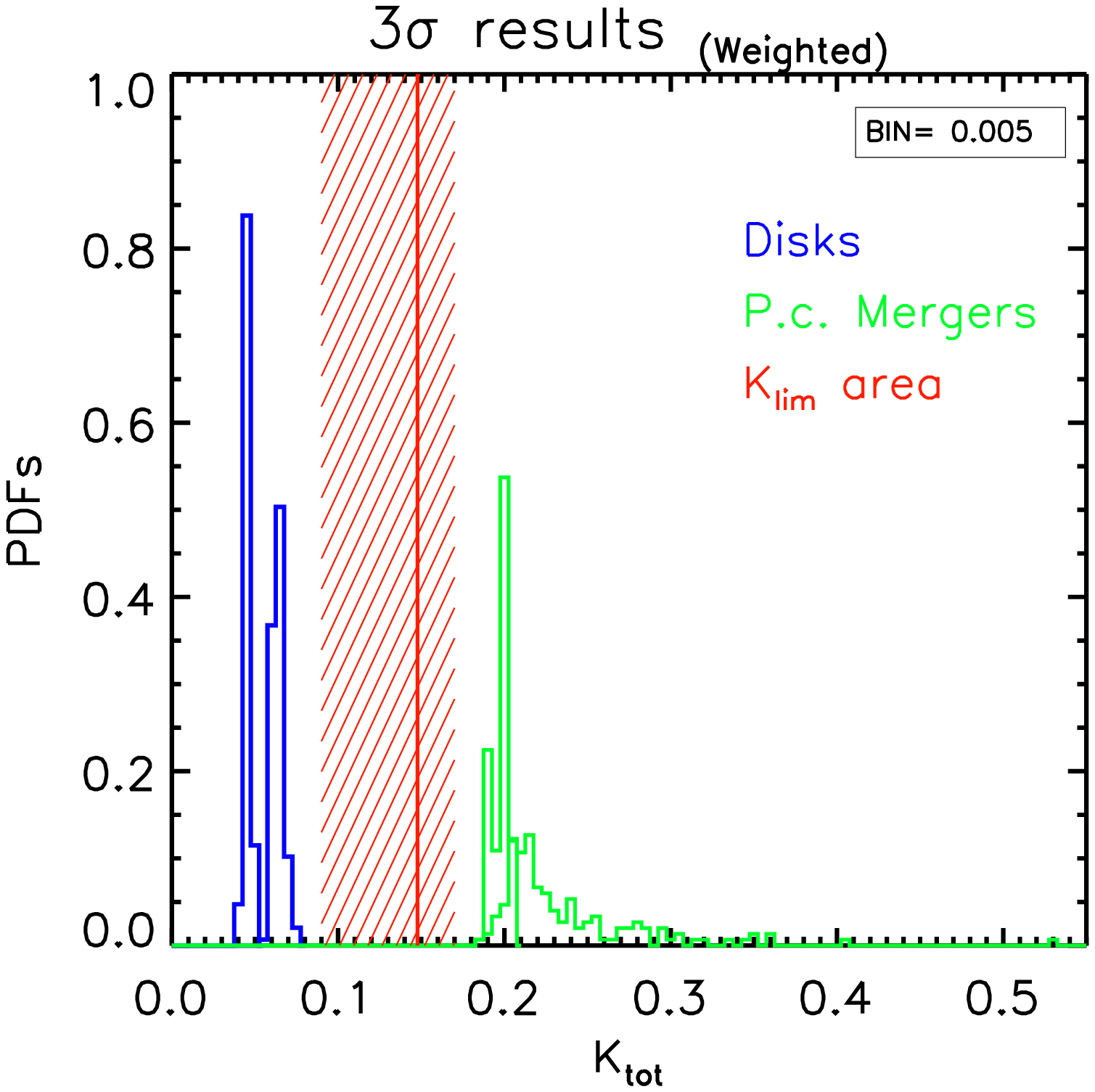} 
\caption{\small {\bf On the top:} Weighted asymmetry measures of the velocity v$_{asym}^{(w)}$ and velocity dispersion $\sigma_{asym}^{(w)}$ fields as derived from the Monte Carlo realizations for the four objects. For each source 150 MC simulations are run but only the $\pm$3$\sigma$ results are shown. The solid red line indicates the empirical division between disks and post-coalescence mergers at $<K_{tot}>$ = 0.146. {\bf On the bottom:} The probability distribution function (PDFs) as derived from MC realizations. The empirical delineation K$_{lim}$ = 0.146 can well separate the two classes along with a large range of other values (i.e., dashed $K_{lim}$ area).}
\label{MC_w}
\end{center}
\end{figure}

\vspace{5mm}

where {\it N} is the total number of radii considered, $C_n$ the value of the circumference for a given ellipse, the different k$_n$ (k$_n^ v$ and k$_n^ \sigma$) are the deviations concerning respectively the velocity field and velocity dispersion maps, while $B_{1}^ v$ is the rotational curves. We will refer to this approach as the {\it `weighted'} method and its associated plane W-[$\sigma_a$ - v$_a$]). In Fig. \ref{MC_w} the results in the W-[$\sigma_a$ - v$_a$] plane for our four galaxies are shown where MC have been performed similarly as above (Sect. 4.3.2).

The results follow the same general trend but the two classes are separated somewhat better than in the unweighted case (Sec. 4.3). Indeed, for class 0 objects the {\it weighted} velocity asymmetries are lower while, for class 2 objects are somewhat higher than in the [$\sigma_a$ - v$_a$] plane. Therefore, it enhances the fact that post-coalescence mergers have larger deviations at larger radii with respect to pure rotational motions while disks have still lower deviations than those obtained using [$\sigma_a$ - v$_a$]. In this case, the total kinematic asymmetry  border, which distinguishes the two disks and the two post-coalescence mergers, is characterized by a mean value of 0.146.

\subsection  {Angular resolution / Redshift dependence}

The angular resolution effects on the distortions of the velocity fields produced by mergers have been discussed by \citet{kron07} on the basis of simulated velocity fields as a function of redshift (i.e., $0<z<1$). They found that for large (Milky Way type) galaxies the distortions are still visible at intermediate redshifts but partially smeared out, while for small galaxies even strong distortions are not visible in the velocity field at z $\approx$ 0.5.

\cite{Gon10}, simulating LBAs at redshift $z\sim$ 2, found that, in general, galaxies at high redshift present smaller values of K$_{asym}$, i.e., they appear more disky than they actually are. The percentage of galaxies classified as mergers drop from $\sim$ 70 \% to $\sim$ 38 \% from $z= 0$ to $z=3$ according to their simulations.

\begin{figure}
\begin{center}
\includegraphics[width=0.48\textwidth]{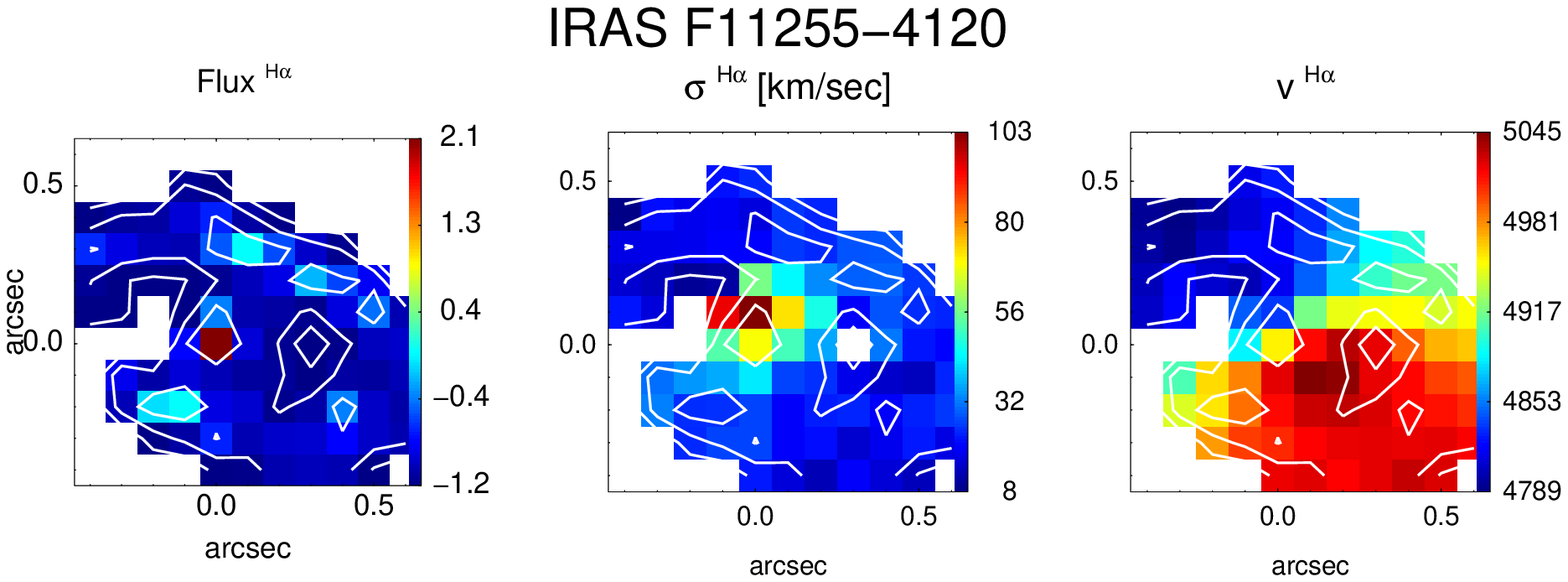}\\
\vspace{2mm}
\includegraphics[width=0.48\textwidth]{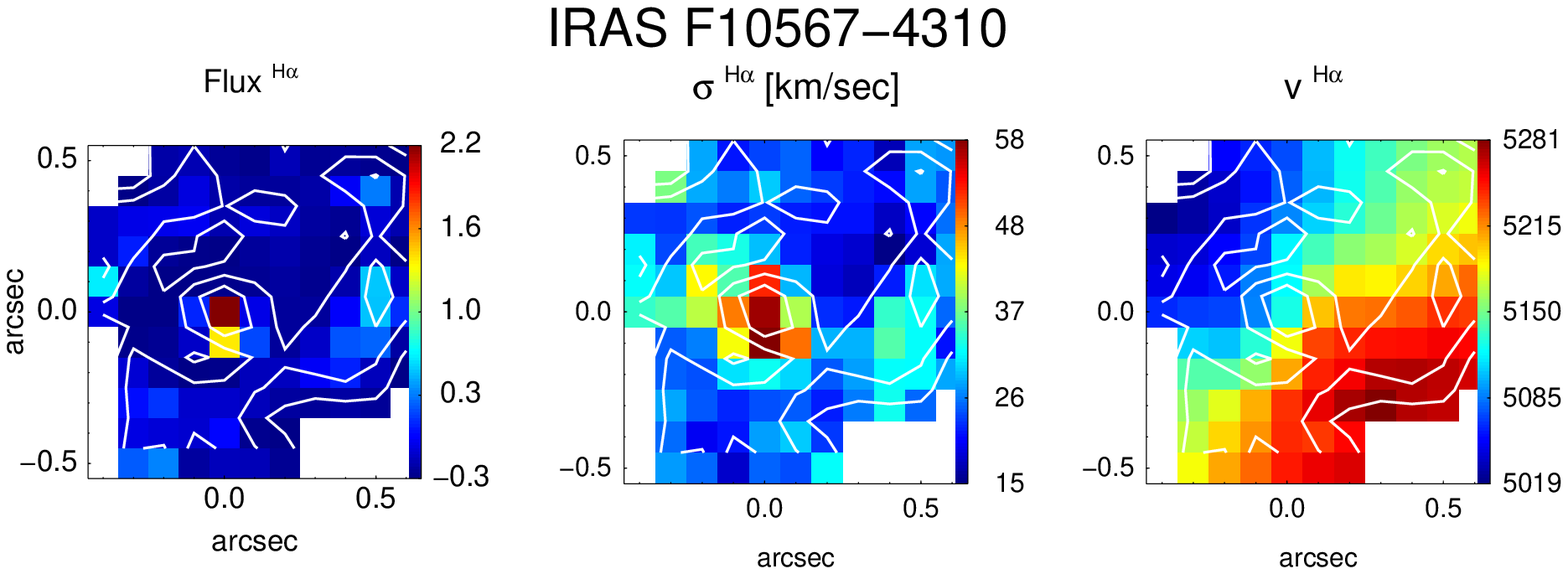}\\
\vspace{2mm}
\includegraphics[width=0.48\textwidth]{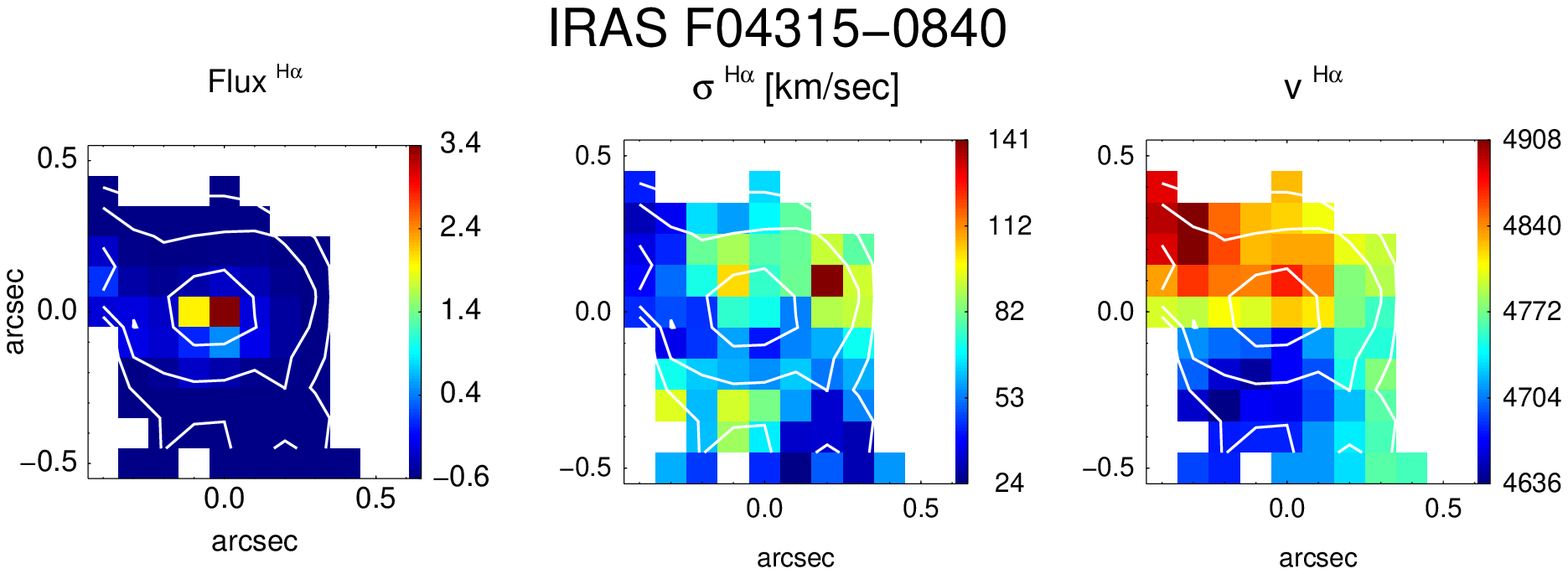}\\
\vspace{2mm}
\includegraphics[width=0.48\textwidth]{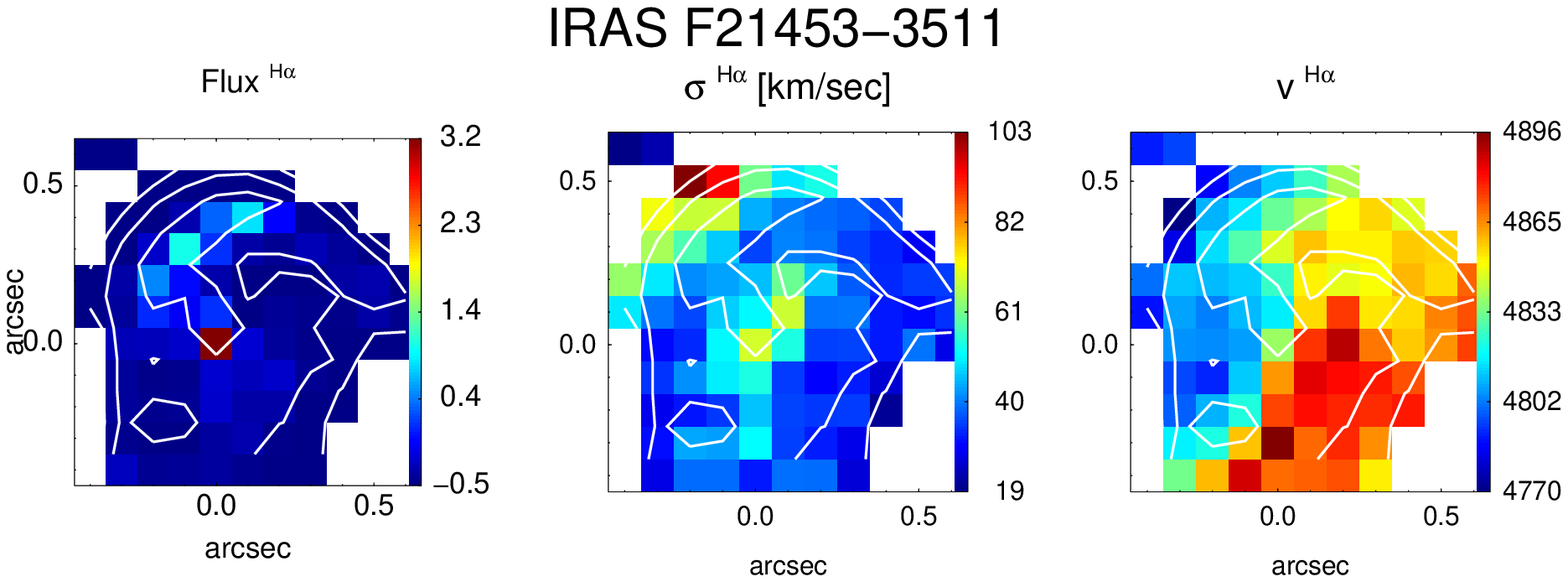}\\
\caption{H$\alpha$ maps for the four objects as observed at $z=3$ with a spatial sampling of 0.1$^{\prime\prime}$. The flux intensity, velocity dispersion $\sigma$ (km/s) and velocity fields $v$ maps (km/s) for the main component are shown. The flux intensity maps are represented in logarithmic scale and in arbitrary flux units. All the images are centered using the H$\alpha$ peak and the iso-contours of the H$\alpha$ flux are over-plotted.}
\label{simulate_maps_highz}\end{center}
\end{figure}

\begin{figure}
\begin{center}
\includegraphics[width=0.4\textwidth]{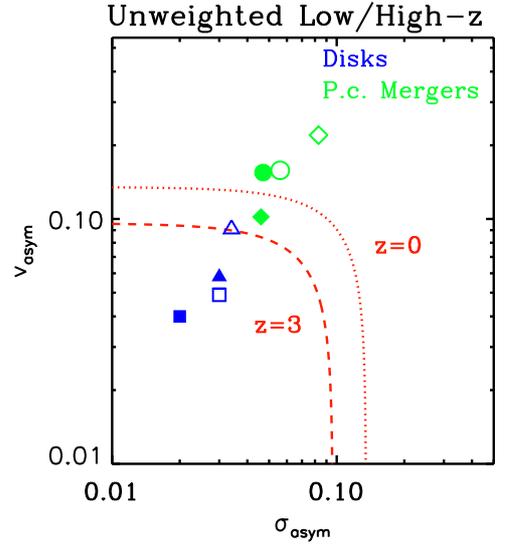} 
\caption{\small The comparison between $local$ and $z=3$ results in the [$\sigma_a$ - v$_a$] plane. Empty symbols represent low-z results while filled ones are for $z=3$. Symbols have the same meaning as in Fig. \ref{Shap_10}. The red dashed line is the {\it high-z} frontier with a value of 0.096 as explained in the text while the dotted one is for the $local$ analysis (i.e., 0.135).} 
\label{altoz}
\end{center}
\end{figure}

\begin{figure}
\begin{center}
\includegraphics[width=0.4\textwidth]{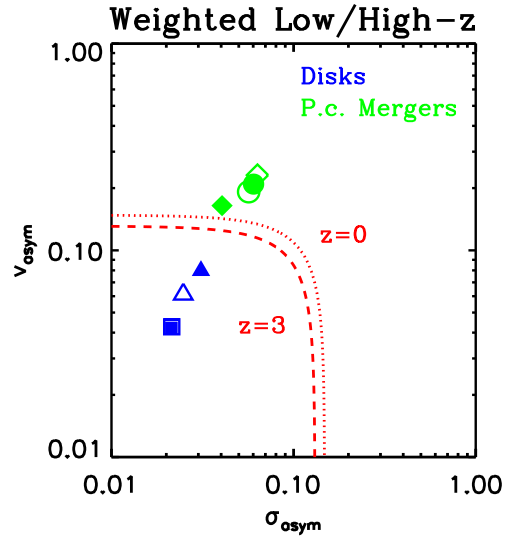}
\caption{\small Comparison between $local$ and $z=3$ results  in the W-[$\sigma_a$ - v$_a$] plane. Empty symbols represent low-z results while filled ones are for z=3. The red dashed line represents the {\it high-z} frontier, while the dotted one is for the local case. } 
\label{ancora}
\end{center}
\end{figure}

In order to investigate the resolution effects on our results we simulate to `observe' these systems at $z=3$ with a typical pixel scale of 0.1$^{\prime\prime}$ (the same pixel scale of the IFU NIRSpec/JWST). At this redshift the current FoV of our images is about 1$^{\prime\prime}$ x 1$^{\prime\prime}$ with a typical scale of 7.83 kpc/arcsec, assuming a $\Lambda$DCM cosmology with H$_0$ = 70 km/s/Mpc, $\Omega_M$ = 0.3 and  $\Omega_\Lambda$ = 0.7. The simulated maps are shown in Fig. \ref{simulate_maps_highz}. We apply $kinemetry$ using these maps and, following the same procedures as before, obtain the results shown in the [$\sigma_a$ - v$_a$] (Fig. \ref{altoz}) and W-[$\sigma_a$ - v$_a$] (Fig. \ref{ancora}) planes. The [$\sigma_a$ - v$_a$] plane shows the expected trend, where both classes appear more symmetric when observed at high redshift. In this case a lower value of the total kinematic asymmetry K$_{asym}$ border is derived (red dashed line), as expected (i.e., K$_{asym}$ = 0.096). Thus, shifting the sample from $z=0$ to $z=3$, the frontier between {\it disks / post-coalescence mergers} in the [$\sigma_a$ - v$_a$] plane changes from 0.135 to 0.096.

The W-[$\sigma_a$ - v$_a$] plane is less sensitive to resolution effects after redshifting our sample at $z = 3$ (see Fig. \ref{ancora}). This is due to the fact that the larger / outer regions are the ones less affected by resolution effects. As with this criterion the associated asymmetries weight more than those present at inner radii, the computed asymmetries (i.e., v$_{asym}$, $\sigma_{asym}$) are less affected by resolution. Therefore, the total kinematic asymmetry distinguishing the two classes changes only from 0.146 to 0.130 between $z= 0$ and $z=3$.

Summarizing, resolution effects tend to smooth kinematic deviations making objects to appear more disky than they actually are. This effects are more significant when analyzing the kinematic asymmetries in the (unweighted) [$\sigma_a$~-v$_a$] plane than in the W-[$\sigma_a$ - v$_a$] one. In particular, when compared our local seeing limited observations with simulated data at $z=3$ (and 0.1$^{\prime\prime}$/spaxel), the total kinematic asymmetry border value is reduced by a 30\% from $z=0$ to $z=3$ in the [$\sigma_a$ - v$_a$] plane, while it is only shifted by 11\% in the W-[$\sigma_a$ - v$_a$] plane.

Comparing our results with those obtained in S08 when considering local spirals and toy-disk models as $observed$ at high redshift, we notice that the low value of our K$_{tot}$ ($\sim$ 0.1) classifies those galaxies as $mergers$ and, on the other hand, the frontier defined in S08 classifies our post-coalescence mergers and IRAS F12112+0305 as $disks$. This illustrates that the definition of $merger$ is a crucial point to define the asymmetry frontier. Indeed, a lower value of the fraction merger/disk (i.e., a higher value of the K$_{tot}$ parameter) can be derived if only pre-coalescence on-going mergers are considered as `true' mergers. The {\it weighted} criterion proposed here  should be applied to larger and more diverse samples in order to understand the uncertainties associated to this type of classifications.
 

\section  {Conclusions}

This paper presents the results from spatially resolved kinematics of four local (z $\sim$ 0.016) luminous infrared galaxies (i.e., LIRGs) observed with the VLT/VIMOS IFU as part of a larger project to characterize the properties of (U)LIRGs on the basis of optical and infrared Integral Field Spectroscopy. The four galaxies are at a similar distance ($\sim$ 70 Mpc) and, on the basis of its morphology, two of them have been classified as {\it isolated disks} and two as post-coalescence $merger$ objects (see Paper III). The velocity field and velocity dispersion maps are analyzed with the aim of studying in detail their kinematics. The $kinemetry$ method (developed by Krajnovi{\'c} and coworkers) is used to characterize the kinematic asymmetries of these galaxies and several criteria are discussed to distinguishing their status. We can draw the following conclusions from this study:

\begin{itemize}

\item
The general kinematic properties of the four LIRGs are consistent with their morphological classification: {\it isolated disks} reveal quite regular velocity fields and centrally peaked velocity dispersion maps consistent with a single rotating disk interpretation while the remaining two show departures from this behavior. In particular we found that post-coalescence {\it mergers} show more irregular velocity fields and velocity dispersion maps showing off-nuclear dispersion peaks (up to 220 km/s at 2.4 kpc from the H$_\alpha$ peak for {IRAS F04315-0840}) or nuclear asymmetric structures;
 
\item
We have found double peaked emission line profiles in the inner regions of the four galaxies. The secondary broad components (i.e., $\sigma\sim $ 70 - 450 km/s) are in all cases blue-shifted ($\Delta v \sim $ 50 - 150 km/s) and, taking into account the large velocities involved, they can be well explained by the presence of an $outflow$ in a dusty environment. The pattern of the 2D kinematic maps of the secondary broad component for the two post-coalescence $mergers$ (with kinematic axes perpendicular to those of the main component) further supports this interpretation. In the particular case of {IRAS F04315-0840} the broad component is found over a quite extended area ($\sim$ 2.4 kpc x 2.7 kpc);

\item	
The v$_c/\sigma_c$ parameter classifies our sources as {\it rotation dominated}. Similar results are obtained when using the quantity v$_{shear}/\Sigma$. This shows that our post-coalescence mergers have a large rotation component and the sole use of these parameters does not allow us to discriminate their kinematic differences with respect to disks;

\item
When the full 2D kinematics information provided by the spatially resolved velocity field and velocity dispersion maps is considered, the kinematic asymmetries are well characterized with $kinemetry$, making the morphological and kinematic classification consistent for the four objects. Disks have lower kinematic asymmetries than those derived for post-coalescence mergers;

\item
We have explored a new criterion to characterize the kinematic asymmetries using {\it kinemetry}. In particular, we introduce a new weighting method which gives weight to the kinematics of the outer regions when computing the total asymmetries v$_{asym}$ and $\sigma_{asym}$. This is motivated by the fact that post-coalescence mergers show relatively small kinematic asymmetries in the inner parts as a consequence of the rapid relaxation into a rotating disk, with the outer parts being still out of equilibrium (i.e., larger asymmetries). The fact that the `frontier' between \textit{disks} and \textit{post-coalescence} systems only changes by 11\% when considering the local and high-z cases suggests that this new criterion is less dependent on angular resolution effects. Thus, the W-[$\sigma_a$ - v$_a$] plane differentiate in a more robust way $disks$ and {\it post-coalescence mergers}.

\item
Classifying $disk/merger$ systems using $kinemetry$ is difficult, and it obviously depends on the definition of $merger$. The `asymmetry frontier' strongly depends on the `type' of mergers considered: if only pre-coalescence on-going mergers are considered as `true' mergers a lower value of the fraction $merger/disk$ systems can be derived. The weighted criterion proposed here helps to characterize in a more robust way post-coalescence asymmetries.

\item
Using previous criteria to classify disks/mergers, our two post-coalescence systems would have been classified as \textit{disks}. This suggests that the ratio of mergers  to disks systems at high-z may have been underestimated. Larger and more diverse samples are required to confirm this conclusion.

\end{itemize}

\newpage

\begin{landscape}
\begin{table}
	\vspace{3cm}
\caption{Kinematic parameters for the sample.}
\vspace{1cm}
\label{pixel}
\begin{scriptsize}
	\begin{tabular}{l cccccccccc } 
\hline\hline\noalign{\smallskip}
Galaxy ID  &   Type of fitting $^{a}$ &$i$ $^{b}$   &{}{  $v_c$ $^{c}$ }   &   {  $\sigma_c$ $^{d}$}  &  { v$_c/ \sigma_c$ }  &{} {  $v_{shear}$ $^{e}$}   & {  $\Sigma$ $^{f}$}    &   { v$_{shear}/ \Sigma$ }  & { $R_{eff}$ } &  { $M_{dyn}$ }  \\
{\smallskip}
{ IRAS code} & {} &{ $degree$} & {   $(km/s)$}   & {   $(km/s)$} & {  }  & 	{   $(km/s)$}   & {   $(km/s)$} & {  }  &    {  $(kpc)$} & {  $(M_\odot)$}  \\
\hline\noalign{\smallskip}
{ {\bf F11255-4120} } & {1c }& { 53$^\circ$ $\pm$ 20$^\circ$} & {   $187 \pm 72$}   & { $104 \pm 3$} & {1.8 $\pm$ 0.7}  & {} & {}  &  {}   &{ } & {  \bf}     \\
{\smallskip}
{ {\bf F11255-4120} } & {2c }& { 53$^\circ$ $\pm$ 20$^\circ$} & {  $188 \pm 76$}   & { $83 \pm 2$} & { 2.3 $\pm$ 1.0}  & 	{125} & {47}  &  {2.7}   &{ $2.40 \pm 0.57$} & {  \bf{$(2.9 \pm  0.8) \cdot 10^{10}$} }     \\
\hline\hline\noalign{\smallskip}
{ {\bf F10567-4310}} &{1c}& { 36$^\circ$ $\pm$ 4$^\circ$} & {  $253 \pm 32$}   & { $56\pm 2$} & { 4.5 $\pm$ 0.8}  & {} & {}  &  {}    & { } &{ \bf }   \\
{\smallskip}
{ {\bf F10567-4310}} & {2c} & { 36$^\circ$ $\pm$ 4$^\circ$} & {  $255 \pm 30$}   & { $ 55.0 \pm 1.4$} & { 4.6 $\pm$ 0.7}  & {120} & {41}  &  {3}  & { $3.20 \pm 0.40$} &{ \bf {$ (1.7 \pm 0.3) \cdot 10^{10} $} }   \\
\hline\hline\noalign{\smallskip}
{ {\bf F04315-0840}} & {1c}& {  n.c.} & {  $157 \pm 3$}   & { $110 \pm 5$} & {1.4 $\pm$ 0.1}  &{} & {}  &  {} & {}  &  {}\\
{\smallskip}
{ {\bf F04315-0840}} & {2c} & {  n.c.} & {  $162 \pm 5$}   & { $69 \pm$ 3} & {2.3 $\pm$ 0.2}  & {51} & {51}  &  {$\sim$ 1}    &	{ $0.94 \pm 0.14$ } & { $(7.9 \pm 1.9) \cdot 10^9 $ }   \\
\hline\noalign{\smallskip}
{ {\bf F04315-0840}} & {1c}& { 29$^\circ\pm3^\circ$} & {  $324 \pm 40$}   & { $110 \pm 5$} & {2.9 $\pm$ 0.5}  &{} & {}  &  {}& {}  &  {} \\
{\smallskip}
{ {\bf F04315-0840}} & {2c}& { 29$^\circ\pm3^\circ$} & {  $335 \pm 45$}   & { $69 \pm 3$} & {4.8 $\pm$ 0.9}  &{51} & {51}  &  {$\sim$ 1}   &{ $0.94 \pm 0.14$ } & { $(7.9 \pm 1.9) \cdot 10^9 $ }\\
\hline\hline\noalign{\smallskip}
{ {\bf F21453-3511}} & {1c} & {  n.c.} & {  $90 \pm 1$}   & { $69 \pm 3$} & {1.30 $\pm$ 0.07}  & {} & {}  &  {}   & { }  & { \bf }     \\		
{\smallskip}
{ {\bf F21453-3511}} &{2c} & { n.c.} & {  $78\pm 2$}   & { $61 \pm 1$} & {1.28 $\pm$ 0.05}  &  {132} & {65}  &  {2}   &{ $ 2.97 \pm 0.82$} & { $(1.9 \pm 0.6) \cdot 10^{10}$ }     \\
\hline\noalign{\smallskip}
{ {\bf F21453-3511}} &{1c} & { 24$^\circ$ $\pm$ 2$^\circ$} & {  $222\pm 21$}   & { $69 \pm 3$} & {3.2 $\pm$ 0.5}  &{} & {}  &  {}    \\
{\smallskip}
{ {\bf F21453-3511}} & {2c} &{ 24$^\circ$ $\pm$ 2$^\circ$} & {  $192 \pm 21$}   & { $61 \pm 1$} & {3.2 $\pm$ 0.4}  &{132} & {65}  &  {2}&{ $ 2.97 \pm 0.82$} & { $(1.9 \pm 0.6) \cdot 10^{10}$ }  \\
\hline\hline\noalign{\smallskip}
	\end{tabular} 
\vskip0.2cm\hskip0.0cm
\end{scriptsize}
\begin{minipage}[h]{21cm}
\tablefoot{ $^{a}$ 1 component fitting (1c) and 2 component fitting (2c, referees to the main (systemic) component of the two-Gaussian fit) is consider for the four galaxies. $^{b}$ Inclination of the galaxy; {\bf n.c.} is when no inclination correction is applied. $^{c}$ Circular velocity derived as the half of the observed peak-to-peak velocity from the H$\alpha$ kinematic. Corrected and no corrected values for the inclination are shown. $^{d}$ Central velocity dispersion as derived from the $\sigma_{H\alpha}$ maps. $^{e}$ Velocity shear not corrected for the inclination of the galaxy. $^{f}$ Global velocity dispersion in the whole galaxy.\\ }
\end{minipage}
\end{table}
\end{landscape}

\begin{acknowledgements}

We acknowledge the anonymous referee for useful comments and suggestions, that helped us to improve the quality of the paper. We also would like to thank Davor Krajnovi{\'c} for his help and valuable comments on his Kinemetry software. This work was funded in part by the Marie Curie Initial Training Network ELIXIR of the European Commission under contract PITN-GA-2008-214227. This work has been supported by the Spanish Ministry of Science and Innovation (MICINN) under grant ESP2007-65475-C02-01. Based on observations carried out at the European Southern observatory, Paranal (Chile), Programs 076.B-0479(A), 078.B-0072(A) and 081.B-0108(A). This research made use of the NASA/IPAC Extragalactic Database (NED), which is operated by the Jet Propulsion Laboratory, California Institute of Technology, under contract with the National Aeronautic and Space Administration.

\end{acknowledgements}

\bibliographystyle{aa} 
\bibliography{biblio}


\begin{appendix} 

\section{ Other kinemetry-based criteria for detecting asymmetries}

In this appendix we describe other criteria that could be useful to distinguish different systems. Indeed, other authors tried to find out criteria able to kinematically classify galaxies at intermediate redshift (i.e., $z \sim 0.6$): for example, \cite{Flores06} developed a simple kinematic classification scheme for distant galaxies based on their 3D kinematics and their morphology in the ACS F775W images. They model the velocity field of a rotating disk that matches the observed velocity gradient to obtain the expected $\sigma$-map corresponding to the observed velocity field and then compare the observed and model $\sigma$-maps, estimating whether the observed kinematics are consistent or not with a rotating disk (i.e., a rotating disk should show a well defined peak in the center of the $\sigma$ map). The spatial separation ($\Delta$r, in pixel) between the peaks in the two $\sigma$-maps and the relative difference ($\epsilon$) between the amplitudes of the modeled and observed $\sigma$ peaks are considered to classify the objects. They found that rotating disks have locations near $\Delta$r $\sim$ $\epsilon$ $\sim$ 0, while galaxies with anomalous velocity fields are clearly offset. The same criterion has been applied in \cite{Yang08}.

Here we propose some criteria, based again on the $kinemetry$ method:

\begin{enumerate}
\item
In this case we consider the normalized higher order deviations $<$k$_5$ / k$_1$$>_v$ of the velocity field with respect to those of the velocity dispersion map $\sigma_{asym}$, since the k$_5$ term indicates complex structure in the velocity map, as noticed in \cite{K06}. Small k$_5$/k$_1$ values should be found in \textit{disk-like} structures while high k$_5$/k$_1$ are expected for mergers, where more complex structures are supposed to exists. This is confirmed by our results: class 0 objects have $<$k$_5$ / k$_1$$>$ close to zero, the amplitude of $k_1$ is substantial, the position angle $\Gamma$ and flattening $q$ remain quite constant such that they can be classified as {\it disk-like} objects. On the other hand, class 2 objects show a rise of the $<$k$_5$ / k$_1$$>$ term up to 0.6 at outer radii, where peculiar structures can be identified. In Fig. \ref{mc_k5}  the results obtained from applying MC simulations (150 for each object). This criterion allows to distinguish the class 0 / 2 in  similar way as in the [$\sigma_a$ - v$_a$] plane, but somewhat worst than in W-[$\sigma_a$ - v$_a$] plane.

\begin{figure}[h]
\begin{center}
\includegraphics[width=0.4\textwidth]{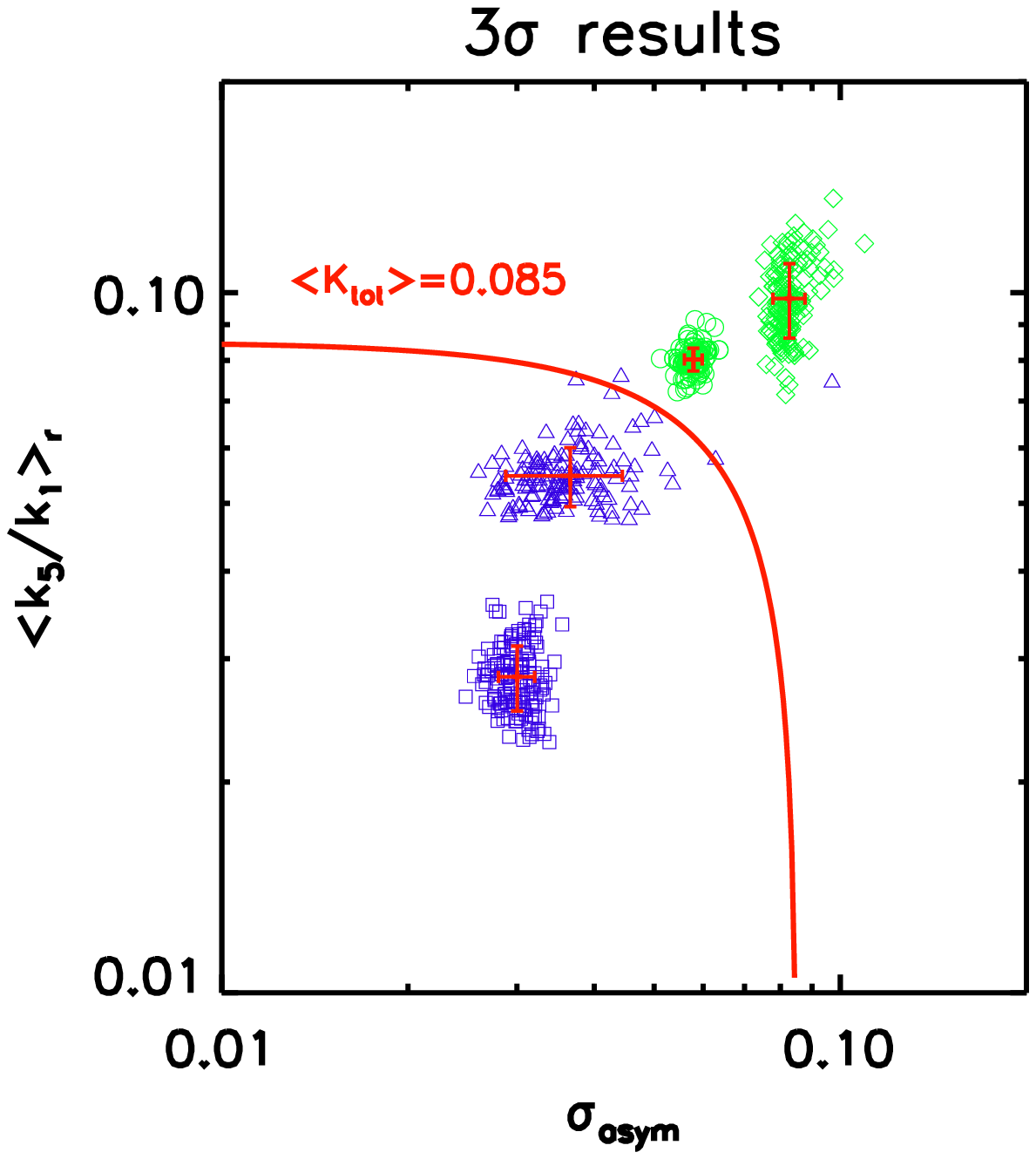}
\includegraphics[width=0.4\textwidth, angle=90]{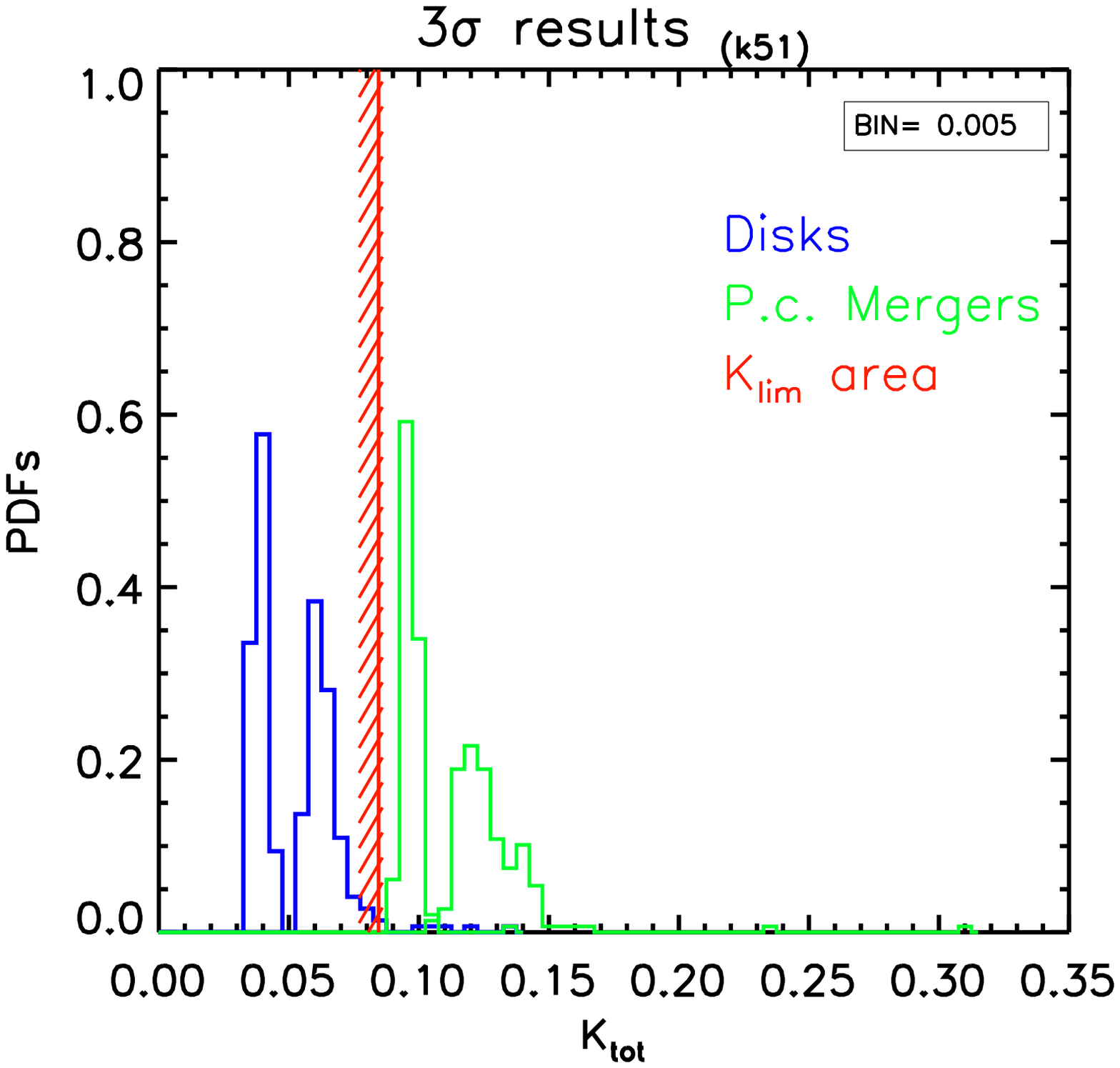} 
\caption{\small {\bf On the top:} Asymmetry measures of the $<$k$_5$/k$_1>$ term and velocity dispersion $\sigma_{asym}^{(w)}$ as derived from the Monte Carlo realizations for the four objects. For each source 150 MC simulations are run but the $\pm$3$\sigma$ results are shown. The solid red line indicates the empirical division between disks and mergers at $<$K$_{lim}>$ = 0.085. {\bf On the bottom:} The probability distribution function (PDFs) as derived from the respective MC realizations. The empirical delineation well separate the two classes defining a small range of possible values (i.e., K$_{lim}$ area). Small deviations $<$k$_5$/k$_1>$ are found for $disks$ while higher values for $mergers$, consistent with our expectation.}
\label{mc_k5}
\end{center}
\end{figure}
	
\vspace{5mm}	
\item 

In some cases the data quality does not allow us to apply sophisticated criteria to look for asymmetries. In these cases a simple criterion may be the only possible option. Therefore, we have studied the potential of the {\it Simple-}[$\sigma_a$ - v$_a$] plane (hereafter, {\it S-}[$\sigma_a$ - v$_a$]), where only the first order correction in the harmonic expansion is considered and asymmetries are defined as  the ratio between the root mean square (RMS) of the residual map and the mean values of the $B_{1,v}$ term over all the radii. The next formula describes the two asymmetries:

$$v_{asym} =  \frac{RMS}{\langle B_{1, v}\rangle}_r  \hspace{1cm} \sigma_{asym} =  \left\langle \frac{ k_{1, \sigma} }  {B_{1, v} }  \right\rangle_r$$

\vspace{3mm}

where the $RMS$ gives an estimate of the deviations of the system from the ideal case of simple rotation in the system\footnote{Residuals are from measured velocities and B$_1$ coefficient.}. In Fig. \ref{Shap_3terms} the results in {\it S -} [$\sigma_a$ - v$_a$] plane are shown. Comparing this plot with that shown in Fig. \ref{Shap_10}, we notice that the RMS/$<$B$_1$$>$ term tends to be higher than the $v_{asym}$ values while the $\sigma_{asym}^{3 terms}$ range smaller than the $\sigma_{asym}^{10 terms}$ range. The $\sigma_{asym}^{3 terms}$ does not seem to characterize well the kinematic asymmetries of these systems (i.e., quite distorted for the {IRAS F04315-0840}). On the other hand, the RMS/$<$B$_1$$>$ term by itself can distinguish the two classes relatively well, where a border dividing them can be estimated with a value of $\sim$ 0.2.

\begin{figure}[h]
\begin{center}
\includegraphics[width=0.4\textwidth]{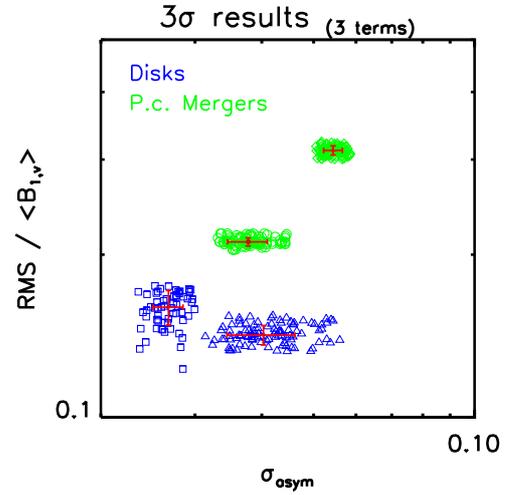}
\caption{\small Subtracting the circular motion (i.e., B$_1$ coefficient) from the data we end up with the velocity residual maps, characterized by mean residuals close to zero and RMS lower than 20 km/s for the whole sample. Having the four objects almost the same RMS, this plot clearly implies higher circular velocities for the class 0 galaxies while lower B$_1$ term is derived for the class 2 ones. Their rotation is well shown in Figs. (\ref{map_rec_11255} - \ref{map_rec_21453}). } 
\label{Shap_3terms}
\end{center}
\end{figure}
 
\end{enumerate}

\end{appendix}

\end{document}